\documentclass[9pt,a4paper]{article}
\usepackage[T1]{fontenc}
\usepackage[utf8]{inputenc}
\usepackage{authblk}
\usepackage{amsmath,amsthm,amssymb}
\usepackage{calligra,indentfirst}
\calligra \newtheorem{theo}{Theorem}[section]
\newtheorem{lem}{Lemma}[section]
\newtheorem{cor}{Corollary}[section]
\newtheorem{prop}{Proposition}[section]

\allowdisplaybreaks
\title{A class of constacyclic codes over $\mathbb{F}_{p^m}[u]/\left<u^2\right>$}
\author[1]{Anuradha Sharma\thanks{Corresponding author, Email address: anuradha@iiitd.ac.in}}
\author[2]{Saroj Rani}
\affil[1]{Department of  Mathematics, IIIT-Delhi, New Delhi-110020, India}
\affil[2]{Department of Mathematics, S.A. Jain College, Ambala, India}

\begin{document}
 \date{} \maketitle
\begin{abstract}
 Let $p$ be an odd prime, and let $m$ be a positive integer satisfying $p^m \equiv 3~(\text{mod }4).$ Let $\mathbb{F}_{p^m}$ be the finite field with $p^m$ elements, and let $R=\mathbb{F}_{p^m}[u]/\left<u^2\right>$ be the finite commutative chain ring with unity. In this paper, we determine all constacyclic codes of length $4p^s$ over $R$ and their dual codes, where $s$ is a positive integer. We also determine their sizes and list some isodual constacyclic codes of length $4p^s$ over $R.$
 \end{abstract}
\section{Introduction}
Berlekamp \cite{berl} first introduced and studied constacyclic codes over finite fields as generalizations of cyclic and negacyclic codes.  These codes have rich algebraic structures and can be easily encoded  and decoded using linear shift registers, which justify their preferred role from engineering perspective.
Recently, it has been observed \cite{convay,hammons,vera,sole,wan}   that many important non-linear codes, such as Kerdock and Preparata codes, are related to linear codes over the ring $\mathbb{Z}_{4}$ of integers modulo 4  with the help of a Gray map.  Since then, codes over finite chain rings have received a lot of attention.  However, the algebraic structure of constacyclic codes over finite chain rings is known only for some special lengths and over certain special finite chain rings. Towards this,  Dinh et al. \cite{dinh} studied the algebraic structures of simple-root cyclic and negacyclic codes  over finite commutative chain rings  and their dual codes. In the same work, they also derived algebraic structures of negacyclic codes of length $2^t$ over the ring $\mathbb{Z}_{2^m}$ of integers modulo $2^m$ and their dual codes, where $t\geq 1$ and  $m \geq 2$ are  integers.  Below we provide a brief overview of some of the recent results known on repeated-root constacyclic codes over finite chain rings.

To describe the recent work, let $p$ be a prime, $s, m$ be  positive integers, and let $\mathbb{F}_{p^m}$ be the finite field of order $p^m.$   Let $R=\mathbb{F}_{p^m}[u]/\left<u^2\right>$ be the finite commutative chain ring with unity. Dinh \cite{dinh4} determined all constacyclic codes of length $p^s$ over  $R$ and their Hamming distances.  Later, Dinh et al. \cite{dinh5} determined algebraic structures of all negacyclic codes of length $2p^s$ over $R$ in terms of their generator polynomials, where $p$ is an odd prime. In a simultaneous work, Liu et al. \cite{liu} determined all cyclic and negacyclic codes of length $2p^s$ over  $R,$ where $p$ is an odd prime.  Chen et al. \cite{chen} determined all constacyclic codes of length $2p^s$ over $R$ and their dual codes, where  $p$ is an odd prime. In a recent work,  Sharma \&\ Rani \cite[Ch. 6]{th} and Dinh et al. \cite{DDS17} simultaneously considered the case $p^m\equiv 1~(\text{mod }4)$ and  determined  all constacyclic codes of length $4p^s$ over $R$ by making the key observation that  any non-zero polynomial of degree less than 4 over $\mathbb F_{p^m}$ is a unit in the ring $\frac{R[x]}{\left< x^{4p^s}-\lambda\right>}$  when $\lambda$ is not a square in $R.$ It is also observed that this is no longer true for the case $p^m\equiv 3 \pmod 4$ (see \cite[Example 6.1]{DDS17}). 

 In this paper, we consider the case $p^m \equiv 3~(\text{mod }4)$ and we establish algebraic structures of all constacyclic codes of length $4p^s$ over $R$   and their dual codes. We  also determine their sizes and list some isodual constacyclic codes of length $4p^s$ over $R.$ This is a follow-up of our  previous work \cite{cn} with Dinh and Sriboonchitta, in which we determined  all cyclic and negacyclic codes of length $4p^s$ over $R$ and their dual codes, and  listed some self-dual cyclic and negacyclic codes of length $4p^s$ over $R.$

This paper is structured as follows: In Section 2, we state some preliminaries. In Section 3, we consider the case $p^m \equiv 3~(\text{mod }4),$ and we determine all $(\alpha+\beta u)$-constacyclic codes of length $4p^s$ over $R$ for all $\alpha (\neq 0),~\beta \in \mathbb{F}_{p^m}$ with $\alpha$ not a square in $\mathbb{F}_{p^m},$   by considering the following two cases separately: \textit{(i)}  $\beta \neq 0$  (Theorem \ref{t11})  and \textit{(ii)} $\beta = 0$  (Lemma \ref{l7} and Theorem \ref{t7}). We also determine their dual codes (Theorems  \ref{d4},  \ref{tt6} and Lemma \ref{d3}) and list some isodual constacyclic codes of length $4p^s$ over $R$ (Corollaries  \ref{c4} and \ref{c3}).
\section{Some preliminaries}
Let $R$ be a finite commutative ring with unity, and let $n$ be a positive integer. An ideal $I$ of $R$ is called a principal ideal if it can be generated by a single element. The ring $R$ is called a \textit{(i)} principal ideal ring if every ideal of $R$ is principal, \textit{(ii)}  local ring if it has a unique maximal ideal (consisting of all the non-units of $R$) and   \textit{(iii)}   a chain ring if all the ideals of $R$ form a chain with respect to the  inclusion relation. Now the following result is well-known:
\begin{prop}\label{pr1}\cite{dinh} Let $R$ be a finite commutative ring with unity. The following statements are equivalent:
\begin{enumerate}\vspace{-1mm}\item[(a)] $R$ is a local ring and the unique maximal ideal $M$ of $R$ is principal, i.e., $M=\left< \gamma\right>$ for some $\gamma \in R.$
\vspace{-1mm}\item[(b)] $R$ is a local principal ideal ring.
\vspace{-1mm}\item[(c)] $R$ is a chain ring whose ideals are given by $\left<\gamma^i\right>,$ $0 \leq i \leq e,$ where $e$ is the nilpotency index of $\gamma.$\end{enumerate} Furthermore, if $R$ is a finite chain ring with the unique maximal ideal as $\left<\gamma \right>$ and the nilpotency index of $\gamma$ is $e,$ then we have $|\left<\gamma^i \right>|=|R/\left<\gamma\right>|^{e-i}$ for $0 \leq i \leq e.$ (Throughout this paper, $|A|$ denotes the cardinality of the set $A.$) \end{prop}
Next let $R^n$ be the $R$-module consisting of all $n$-tuples over $R.$ For a unit $\lambda \in R,$ a $\lambda$-constacyclic code $\mathcal{C}$ of length $n$ over $R$ is defined as an $R$-submodule of $R^n$ satisfying the following property: $(c_0,c_1,\cdots,c_{n-1})\in \mathcal{C}$ implies that $(\lambda c_{n-1},c_0,c_1,\cdots,c_{n-2}) \in \mathcal{C}.$ The dual code of $\mathcal{C},$ denoted by $\mathcal{C}^{\perp},$ is defined as $\mathcal{C}^{\perp}=\{a \in R^n: a.c=0 \text{ for all }c \in \mathcal{C}\},$ where $a.c=a_0c_0+a_1c_1+\cdots + a_{n-1}c_{n-1}$ for $a=(a_0,a_1,\cdots,a_{n-1}) \in R^n$ and $c=(c_0,c_1,\cdots,c_{n-1}) \in \mathcal{C}.$  It is easy to observe that $\mathcal{C}^{\perp}$ is a $\lambda^{-1}$-constacyclic code of length $n$ over $R.$ The code $\mathcal{C}$ is said to be  isodual if the codes $\mathcal{C}$ and $\mathcal{C}^{\perp}$ are $R$-linearly  equivalent, i.e., if there exists a monomial transformation $T: \mathcal{C} \rightarrow \mathcal{C}^{\perp}$ defined as $T(c_0,c_1,\cdots,c_{n-1})=(u_0 c_{\sigma(0)},u_1 c_{\sigma(1)},\cdots,u_{n-1}c_{\sigma(n-1)}) $ for all $(c_0,c_1,\cdots,c_{n-1}) \in \mathcal{C},$ where $\sigma$ is a permutation of $\{0,1,2,\cdots,n-1\}$ and $u_0,u_1,\cdots,u_{n-1}$ are units in $R.$

Under the standard $R$-module isomorphism $\phi: R^n \rightarrow R[x]/\left<x^n-\lambda\right>,$ defined as $\phi(a_0,a_1,\cdots,a_{n-1})=a_0+a_1 x+\cdots+a_{n-1}x^{n-1}$ for each $(a_0,a_1,\cdots,a_{n-1})\in R^n,$ the code $\mathcal{C}$ can be identified as an ideal of the ring $R[x]/\left<x^n-\lambda\right>.$ Thus the study of $\lambda$-constacyclic codes of length $n$ over $R$ is equivalent to the study of ideals of the ring $R[x]/\left<x^n-\lambda\right>.$ Further, the dual code $\mathcal{C}^{\perp}$ of $\mathcal{C}$  is given by $\mathcal{C}^{\perp}=\{u(x) \in  R[x]/\left<x^n-\lambda^{-1}\right>: u(x)c^*(x)=0 \text{ in } R[x]/\left<x^n-\lambda^{-1}\right> \text{ for all } c(x) \in \mathcal{C}\},$ where $c^*(x)=x^{\text{deg }c(x)}c(x^{-1})$ for all $c(x) \in \mathcal{C}\setminus\{0\}$ and $c^*(x)=0$ if $c(x)=0.$ The annihilator of $\mathcal{C}$ is defined as $\text{ann}(\mathcal{C})=\{f(x) \in R[x]/\left<x^n-\lambda\right>: f(x) c(x)=0 \text{ in }R[x]/\left<x^n-\lambda\right> \text{ for all } c(x) \in \mathcal{C}\}.$ It is easy to observe that $\text{ann}(\mathcal{C})$ is an ideal of $R[x]/\left<x^n-\lambda\right>.$ Further, for any ideal $I$ of $R[x]/\left<x^n-\lambda \right>,$ we define $I^*=\{f^*(x): f(x) \in I\},$ where $f^*(x)= x^{\text{deg }f(x)}f(x^{-1})$ if $f(x) \neq 0$ and $f^*(x)=0$ if $f(x)=0.$	 It is easy to see that $I^*$ is an ideal of the ring $R[x]/\left<x^n-\lambda^{-1}\right>.$ Then the following holds.
\begin{lem}\cite{chen} \label{pr2}If $\mathcal{C} \subseteq R[x]/\left<x^n-\lambda\right>$ is a $\lambda$-constacyclic code of length $n$ over $R,$ then we have \vspace{-1mm}\begin{equation*}\mathcal{C}^{\perp}=\text{ann}(\mathcal{C})^*.\vspace{-1mm}\end{equation*} \end{lem}
The following two lemmas are quite useful in the determination  of dual codes of $\lambda$-constacyclic codes of length $n$ over $R.$
\noindent\begin{lem}\label{pr3} \cite{chen} For $f(x), g(x) \in R[x]/\left<x^n-\lambda \right>,$ define $(f+g)(x)=f(x)+g(x)$ and $(fg)(x)=f(x)g(x).$ \begin{enumerate}
\item[(a)] If $f(x),g(x), (f+g)(x)$ all are non-zero and $\text{deg }f(x) \geq \text{deg }g(x),$ then we have \vspace{-1mm}\begin{equation*}\left<(f+g)^*(x)\right>=\left<f^*(x)+x^{\text{deg }f(x)-\text{deg }g(x)}g^*(x)\right>.\vspace{-1mm}\end{equation*} \vspace{-6mm}\item[(b)] If $f(x),g(x),(fg)(x)$ all are non-zero, then $\left<(fg)^*(x)\right>=\left<f^*(x)g^*(x)\right>.$ \end{enumerate}\end{lem}
\noindent \begin{lem}\cite{chen}\label{pr4} Let $I=\left<f(x),ug(x)\right>$ be an ideal of $R[x]/\left<x^n-\lambda \right>.$ We have $$I^*=\{h^*(x): h(x) \in I\}=\left<f^*(x),ug^*(x)\right>.$$\end{lem}
Throughout this paper, let $\mathbb{F}_{p^m}$ be the finite field of order $p^m,$ where $p$ is an odd prime and $m$ is a positive integer satisfying $p^m \equiv 3~(\text{mod }4).$ Let $R=\mathbb{F}_{p^m}[u]/\left<u^2\right>$ be the finite commutative chain ring with unity and having the unique maximal ideal as $\left<u\right>.$ Note that each element $\lambda \in R$ can be uniquely expressed as $\lambda=\alpha+\beta u,$ where $\alpha,\beta \in \mathbb{F}_{p^m}.$
\begin{lem} \cite{chen}\label{pr5} Let $\lambda  =\alpha+\beta u \in R,$ where $\alpha,\beta \in \mathbb{F}_{p^m}.$  Then the following hold. \begin{enumerate}\item[(a)] $\lambda$ is a unit in $R$ if and only if $\alpha \neq 0.$ \item[(b)] $\lambda$ is a square in $R$ if and only if $\alpha$ is a square in $\mathbb{F}_{p^m}.$
\item[(c)] There exists $\alpha_0 \in \mathbb{F}_{p^m}$ satisfying $\alpha_0^{p^s}=\alpha.$ \end{enumerate}\end{lem}
 \noindent For a code $\mathcal{C}$ of length $n$ over $R,$ the torsion and residue codes of $\mathcal{C}$ are codes of length $n$ over $\mathbb{F}_{p^m}$ defined as follows:
 $$\text{Tor}(\mathcal{C})=\{ a \in \mathbb{F}_{p^m}^n: ua \in \mathcal{C}\}$$ and $$ \text{Res}(\mathcal{C})=\{a \in \mathbb{F}_{p^m}^n: a+ub \in \mathcal{C} \text{ for some }b \in \mathbb{F}_{p^m}^n\}.$$
 \begin{lem} \cite{chen}\label{pr6} If $\mathcal{C}$ is an $(\alpha+\beta u)$-constacyclic code of length $n$ over $R,$ then  both $\text{Tor}(\mathcal{C})$ and $\text{Res}(\mathcal{C})$ are $\alpha$-constacyclic codes of length $n$ over $\mathbb{F}_{p^m}.$ Moreover, we have $|\mathcal{C}|=|\text{Tor}(\mathcal{C})||\text{Res}(\mathcal{C})|.$\end{lem}
 From now onwards, we shall focus our attention on constacyclic codes of length $4p^s$ over the ring $R=\mathbb{F}_{p^m}[u]/\left<u^2\right>,$  where $p$ is an odd prime and $s,m$ are positive integers satisfying $p^m \equiv 3~(\text{mod }4).$
 \section{Algebraic structures of constacyclic codes of length $4p^s$ over $R$}
 In this section, we shall establish algebraic structures of all $\lambda$-constacyclic codes of length $4p^s$ over $R,$ where $\lambda$ is a unit in $R$ and $s$ is a positive integer. 

If $\lambda$ is a square in $R,$ say $\lambda=\lambda_{0}^2$ for some $\lambda_0 \in R,$ then by Chinese Remainder Theorem, we have $$\frac{R[x]}{\left<x^{4p^s}-\lambda\right> }\simeq \frac{R[x]}{\left<x^{2p^s}-\lambda_{0}\right>} \oplus \frac{R[x]}{\left<x^{2p^s}+\lambda_{0}\right>}.$$ From this, it follows that every $\lambda$-constacyclic code of length $4p^s$ over $R$ is a direct sum of a $\lambda_{0}$-constacyclic  code of length $2p^s$ over $R$ and a $(-\lambda_{0})$-constacyclic code of length $2p^s$ over $R.$ Note that the algebraic structures of all constacyclic codes of length $2p^s$ over $R$ and their dual codes have been determined in \cite{chen}. In view of this, henceforth, we assume that $\lambda$ is not a square in $R.$ Let us write $\lambda=\alpha+\beta u,$ where $\alpha, \beta \in \mathbb{F}_{p^m}.$
By Lemma \ref{pr5}(a) and (b), we note that $\alpha \neq 0$ and $\alpha$ is not a square in $\mathbb{F}_{p^m}.$ Now we shall distinguish the following two cases: \textit{(i)} $\beta \neq 0$  and \textit{(ii)} $\beta=0.$

\subsection{The case $\beta \neq 0$}
In this section, we shall determine algebraic structures of all $(\alpha+\beta u)$-constacyclic codes of length $4p^s$ over $R,$ where
  both $\alpha,\beta $ are non-zero elements of $\mathbb{F}_{p^m}$ and $\alpha$ is not a square in $\mathbb{F}_{p^m}.$ For this, we recall that each $(\alpha+\beta u)$-constacyclic code of length $4p^s$ over $R$ is an ideal of the quotient ring $\mathcal{R}_{\alpha,\beta}=R[x]/\left<x^{4p^s}-(\alpha+\beta u)\right>.$ By Lemma \ref{pr5}(c), we see that there exists $\alpha_0 \in \mathbb{F}_{p^m}$ satisfying $\alpha_0^{p^s}=\alpha.$ Then we observe the following:
\begin{prop}\label{p1} \begin{enumerate}
\item[(a)] In $\mathcal{R}_{\alpha,\beta},$ we have $\left<(x^4-\alpha_0)^{p^s}\right>=\left<u\right>.$ As a consequence, $x^4-\alpha_0$ is a nilpotent element of $\mathcal{R}_{\alpha,\beta}$ and its nilpotency index is $2p^s.$
\item[(b)] The equation $x^4+ 4 \alpha_0=0$ has exactly two roots in $\mathbb{F}_{p^m}.$  In fact, if $\gamma$ is a root of $x^4+4 \alpha_0=0,$ then $-\gamma$ is its other root. \item[(c)] We have $x^4-\alpha_0=\left(x^2+\gamma x+\frac{\gamma^2}{2}\right)\left(x^2-\gamma x+\frac{\gamma^2}{2}\right).$ Furthermore, the polynomials $x^2+ \gamma x +\frac{\gamma^2}{2}$ and $x^2- \gamma x +\frac{\gamma^2}{2}$ are coprime irreducible polynomials over $R.$
 \end{enumerate} \end{prop}
\noindent \textbf{Proof.}  Proof is trivial. $\hfill \Box$
\begin{prop} \label{p2}Let $p(x)=ax^3+bx^2+cx+d$ be a non-zero polynomial in $\mathbb{F}_{p^m}[x].$  Then
 $p(x)$ is a unit in $\mathcal{R}_{\alpha,\beta}$ if and only if \vspace{-1mm}\begin{equation*} p(x) \not\in \left<x^2+\gamma x+\frac{\gamma^2}{2}\right> \cup \left<x^2-\gamma x+\frac{\gamma^2}{2}\right>,\vspace{-1mm}\end{equation*}   where $\gamma \in \mathbb{F}_{p^m}$ satisfies $\gamma^4+4\alpha_0=0.$
 As a consequence,  we have \vspace{-1mm}\begin{equation*} \left<x^4-\alpha_0\right> \subsetneq \left<x^2+\gamma x+\frac{\gamma^2}{2}\right> \text{and } \left<x^4-\alpha_0\right> \subsetneq \left< x^2-\gamma x+\frac{\gamma^2}{2}\right>  \text{ in } \mathcal{R}_{\alpha,\beta}. \vspace{-1mm}\end{equation*}
\end{prop}
 \noindent\textbf{Proof.} When $a=b=c=0,$ we have $p(x)=d (\neq 0),$ which is trivially a unit in $\mathcal{R}_{\alpha,\beta}.$ Now we assume that $a,b,c$ are not all zero, and  we shall consider the following  cases separately: \textbf{(i)} $a=b=0$ and $ c \neq 0,$ \textbf{(ii)} $a=c=0$ and $b \neq 0,$  \textbf{(iii)} $a=d=0$ and both $b ,c$ are non-zero, \textbf{(iv)} $a \neq 0$ and $b=c=d=0,$ \textbf{(v)} $a=0$ and $b,c,d$ all are non-zero,  \textbf{(vi)} $a \neq 0,$  $d=0$ and $b,c$ are not both zero \textbf{(vii)} both $a ,d$ are non-zero.
\begin{description}\item[(i)] Let $a=b=0$ and $c \neq 0.$ Then we have $p(x)=cx+d.$  As $\alpha$ is not a square in $\mathbb{F}_{p^m},$ we note that $\alpha- (c^{-1}d)^{4p^s} \neq 0,$ which, by Lemma \ref{pr5}(a),  implies that $\alpha+\beta u-(c^{-1}d)^{4p^s}$ is a unit in $R.$ Further, it is easy to see that  \vspace{-1mm}\begin{equation*} p(x)c^{-1}(x+c^{-1}d)^{p^s-1} (x-c^{-1}d)^{p^s}(x^2+c^{-2}d^2)^{p^s}(\alpha+\beta u-(c^{-1}d)^{4p^s})^{-1}=1\text{ in }\mathcal{R}_{\alpha,\beta}.\vspace{-1mm}\end{equation*}   From this, it follows that $p(x)$ is a unit in $\mathcal{R}_{\alpha,\beta}.$
\item[(ii)] Let $a=c=0$ and $b \neq 0.$ Here $p(x)=bx^2+d.$ If $d=0,$ then by case \textbf{(i)}, we see that $p(x)=bx^2=(bx)x$ is a unit in $\mathcal{R}_{\alpha,\beta}.$ When $d$ is non-zero, we have $p(x)=bx^2+d=b(x^2+b^{-1}d).$  Since $\alpha$ is not a square in $\mathbb{F}_{p^m},$  we see that  $\alpha-(b^{-1}d)^{2p^s} \neq 0.$ Now by Lemma \ref{pr5}(a),  we note that $\alpha-(b^{-1}d)^{2p^s}+\beta u$ is a unit in $R.$ Further, it is easy to verify that \vspace{-1mm}\begin{equation*}
 p(x) b^{-1}(x^2+b^{-1}d)^{p^s-1}(x^2-b^{-1}d)^{p^s} \{\alpha-(b^{-1}d)^{2p^s}+\beta u\}^{-1}=1 \text{ in }\mathcal{R}_{\alpha, \beta}.\vspace{-1mm}\end{equation*} 
This implies that $p(x)$ is a unit in $\mathcal{R}_{\alpha,\beta}.$
\item[(iii)] When $a=d=0$ and both $b ,c$ are non-zero, we have $p(x)=bx^2+cx=x(bx+c),$ which is a unit in $\mathcal{R}_{\alpha,\beta}$ by case \textbf{(i)}. 
\item[(iv)] When $a \neq 0$ and $b=c=d=0,$ we have $p(x)=ax^3,$ which is a unit in $\mathcal{R}_{\alpha,\beta}$ by case \textbf{(i)}.
\item[(v)]   Suppose that $a=0$ and $b,c,d$ all are non-zero.  Here we have \vspace{-1mm}\begin{equation*}
p(x)=bx^2+cx+d=b(x^2+b^{-1}cx+b^{-1}d)=b(x^2+c_1x+d_1),\vspace{-1mm}\end{equation*} where $c_1=b^{-1}c$ and $d_1=b^{-1}d.$ First we observe that \vspace{-1mm}\begin{equation*}
\alpha_0+(c_1^3-2c_1d_1)x+(d_1c_1^2-d_1^2) \vspace{-1mm}\end{equation*} is a unit in $\mathcal{R}_{\alpha,\beta}$ if and only if \vspace{-1mm}\begin{equation*} \alpha+\beta u +(c_1^3-2c_1d_1)^{p^s}x^{p^s}+(d_1c_1^2-d_1^2)^{p^s}\vspace{-1mm}\end{equation*} is  a unit in $\mathcal{R}_{\alpha,\beta}$  and  
\vspace{-1mm}\begin{equation*} p(x)b^{-1}(x^2+c_1x+d_1)^{p^s-1}(x^2-c_1x-d_1+c_1^2)^{p^s}\{\alpha+\beta u +(c_1^3-2c_1d_1)^{p^s}x^{p^s}+(d_1c_1^2-d_1^2)^{p^s}\}^{-1}=1\vspace{-1mm}\end{equation*} in $\mathcal{R}_{\alpha, \beta}.$ This shows that $\alpha_0+(c_1^3-2c_1d_1)x+(d_1c_1^2-d_1^2)$ is a unit in $\mathcal{R}_{\alpha,\beta}$ if and only if $p(x)$ is a unit in $\mathcal{R}_{\alpha, \beta}.$
Next we observe that $\alpha_0+(c_1^3-2c_1d_1)x+(d_1c_1^2-d_1^2)$ is a non-unit in $\mathcal{R}_{\alpha,\beta}$ if and only if  $c_1 \in \mathbb{F}_{p^m}$ satisfies $c_1^4=-4\alpha_0$ and $d_1=\frac{c_1^2}{2}.$ Now since $u$ is nilpotent in $\mathcal{R}_{\alpha,\beta}$ and $\alpha_0^{p^s}=\alpha,$ we see that $\alpha+\beta u +(c_1^3-2c_1d_1)^{p^s}x^{p^s}+(d_1c_1^2-d_1^2)^{p^s}$ is a non-unit in $\mathcal{R}_{\alpha,\beta}$ if and only if $c_1^4=-4\alpha_0$ and $d_1=\frac{c_1^2}{2}.$  This implies that $x^2+c_1 x+d_1$ is a non-unit in $\mathcal{R}_{\alpha,\beta}$ if and only if  $c_1 \in \mathbb{F}_{p^m}$ satisfies $c_1^4=-4\alpha_0$ and $d_1=\frac{c_1^2}{2}.$  That is, $p(x)=bx^2+cx+d=b(x^2+c_1x+d_1)$ is a non-unit in $\mathcal{R}_{\alpha,\beta}$ if and only if $c_1^4=-4\alpha_0$ and $d_1=\frac{c_1^2}{2}.$ In view of Proposition \ref{p1}(b), we see that $p(x)$ is a non-unit in $\mathcal{R}_{\alpha,\beta}$ if and only if $p(x)$ is either $b(x^2+\gamma x+\frac{\gamma^2}{2})$ or $b(x^2-\gamma x+\frac{\gamma^2}{2}),$ where $\gamma^4+4\alpha_0=0.$
\item[(vi)] Suppose that $a \neq 0,$   $d=0,$ and $b,c$ are not both zero. Here we have \vspace{-1mm}\begin{equation*}
 p(x)=ax^3+bx^2+cx=x(ax^2+bx+c),\vspace{-1mm}\end{equation*} which is a unit in $\mathcal{R}_{\alpha,\beta} $ if and only if \vspace{-1mm}\begin{equation*}(ax^2+bx+c)\not\in \left<x^2+\gamma x+\frac{\gamma^2}{2}\right>\cup \left<x^2-\gamma x+\frac{\gamma^2}{2}\right>\vspace{-1mm}\end{equation*}  by  case \textbf{(v)}. That is,\vspace{-1mm}\begin{equation*}
 p(x)=ax^3+bx^2+cx\vspace{-1mm}\end{equation*}  is a unit in $\mathcal{R}_{\alpha,\beta}$ if and only if \vspace{-1mm}\begin{equation*}
p(x) \not\in \left<x^2+\gamma x+\frac{\gamma^2}{2}\right>\cup \left<x^2-\gamma x+\frac{\gamma^2}{2}\right>.\vspace{-1mm}\end{equation*}
\item[(vii)] Suppose that both $a,d$ are non-zero.  Here we have \vspace{-1mm}\begin{equation*}
p(x)=a(x^3+a^{-1}bx^2+a^{-1}cx+a^{-1}d)=a(x^3+b'_1x^2+c'_1x+d'_1),\vspace{-1mm}\end{equation*} where $b'_1=a^{-1}b,$ $c'_1=a^{-1}c$ and $d'_1=a^{-1}d.$ First we assert that\vspace{-1mm}\begin{equation*}
 p_1(x)=(c'_1-{b'_1}^2)x^2+(d'_1-b'_1c'_1)x+\alpha_0-b'_1d'_1\vspace{-1mm}\end{equation*} is a non-zero polynomial over $\mathbb{F}_{p^m}.$ For, if $c'_1-{b'_1}^2=d'_1-b'_1c'_1=\alpha_0-b'_1d'_1=0,$ then we get $\alpha_0={b'_1}^4.$ This is a contradiction, as $\alpha_0$ is a non-square in $\mathbb{F}_{p^m}.$

Further, it is easy to observe that $p_1(x)$ is a unit in $\mathcal{R}_{\alpha, \beta}$ if and only if  \vspace{-1mm}\begin{equation*}
(\alpha_0+(c'_1-{b'_1}^2)x^2+(d'_1-b'_1c'_1)x-b'_1d'_1)^{p^s}+\beta u \text{ ~is a unit in~ }\mathcal{R}_{\alpha,\beta} \text{ ~and }\vspace{-1mm}\end{equation*}   \vspace{-1mm}\vspace{-2mm}\begin{equation*}
p(x)a^{-1}(x^3+b'_1x^2+c'_1x+d'_1)^{p^s-1}(x-b'_1)^{p^s}\{(\alpha_0+(c'_1-{b'_1}^2)x^2+(d'_1-b'_1c'_1)x-b'_1d'_1)^{p^s}+\beta u \}^{-1}=1\vspace{-1mm}\end{equation*} in $\mathcal{R}_{\alpha, \beta}.$ Next we assert that $p_1(x)(\neq 0)$ is a non-unit in $\mathcal{R}_{\alpha, \beta}$ if and only if \vspace{-1mm}\begin{equation*} p(x) \in  \left<x^2+ \gamma x+\frac{\gamma^2}{2}\right> \cup \left<x^2- \gamma x+\frac{\gamma^2}{2}\right>.\vspace{-1mm}\end{equation*}

For this, we see that if $c'_1-{b'_1}^2=0,$ then $p_1(x)=(d'_1-b'_1c'_1)x+\alpha_0-b'_1d'_1$ is a non-zero polynomial over $\mathbb{F}_{p^m},$ which is a unit in $\mathcal{R}_{\alpha,\beta}$ by case \textbf{(i)}. So we assume that $c'_1-{b'_1}^2 \neq 0.$ If $b'_1=0,$ then we have $p_1(x)=c'_1x^2+d'_1x+\alpha_0,$ which, by case \textbf{(v)}, is a non-unit in $\mathcal{R}_{\alpha,\beta}$ if and only if $d'_1{c'_1}^{-1}=\epsilon \gamma$ and
$\alpha_0{c'_1}^{-1}=\frac{\gamma^2}{2},$ where $\gamma^4=-4\alpha_0$ and $\epsilon \in \{1,-1\}.$ This holds if and only if $d'_1=\epsilon  \gamma c'_1,$ $\alpha_0=\frac{\gamma^2 c'_1}{2}$ and $\gamma^4=-4\alpha_0,$ which implies that $\gamma^4+2\gamma^2 c'_1=0.$  This gives $c'_1=-\frac{\gamma^2}{2}$ and $d'_1=-\frac{\epsilon \gamma^3}{2}.$ From this, it follows  that $p_1(x)$ is a non-unit in $\mathcal{R}_{\alpha, \beta}$ if and only if \vspace{-1mm}\begin{equation*} p(x)=x^3+c'_1 x+d'_1=x^3-\frac{\gamma^2}{2}x-\frac{\epsilon \gamma^3}{2}=\left(x^2+\epsilon \gamma x+\frac{\gamma^2}{2}\right)(x-\epsilon \gamma ),\vspace{-1mm}\end{equation*} which lies in $\left<x^2+\epsilon \gamma x+\frac{\gamma^2}{2}\right>.$ Finally, let $b'_1 \neq 0 $ and $c'_1-{b'_1}^2 \neq 0.$ Then $p_1(x)$ is a non-unit in $\mathcal{R}_{\alpha,\beta}$ if and only if \vspace{-1mm}\begin{equation*} x^2+\left(\frac{d'_1-b'_1c'_1}{c'_1-{b'_1}^2}\right)x+\left(\frac{\alpha_0-b'_1d'_1}{c'_1-{b'_1}^2}\right)\vspace{-1mm}\end{equation*} is a non-unit in $\mathcal{R}_{\alpha,\beta},$ which holds if and only if $\frac{d'_1-b'_1c'_1}{c'_1-{b'_1}^2}=\epsilon \gamma$ and $\frac{\alpha_0-b'_1d'_1}{c'_1-{b'_1}^2}=\frac{\gamma^2}{2}$ by case \textbf{(v)},  where $\epsilon \in \{1,-1\}$ and $\gamma^4=-4\alpha_0.$ This holds if and only if  \vspace{-1mm}\begin{equation*}d'_1=b'_1c'_1+\epsilon \gamma(c'_1-{b'_1}^2) \text{ and } d'_1=\frac{2\alpha_0-\gamma^2(c'_1-{b'_1}^2)}{2b_1},\vspace{-1mm}\end{equation*}  which gives \vspace{-1mm}\begin{equation*} c'_1=\frac{2\alpha_0+{b'_1}^2\gamma^2+2{b'_1}^3\epsilon \gamma}{2{b'_1}^2+2b'_1\epsilon \gamma +\gamma^2} \text{ and } d'_1=\frac{2\alpha_0 b'_1+2\alpha_0 \epsilon \gamma +{b'_1}^3\gamma^2}{2{b'_1}^2+2b'_1\epsilon \gamma +\gamma^2}.\vspace{-1mm}\end{equation*}  From this and using the fact that $\gamma^4=-4\alpha_0,$ we see that $p_1(x)$ is a non-unit in $\mathcal{R}_{\alpha, \beta}$ if and only if
\begin{eqnarray*} p(x)&=&x^3+b'_1x^2+c'_1x+d'_1=x^3+b'_1x^2+\left(\frac{2\alpha_0+{b'_1}^2\gamma^2+2{b'_1}^3\epsilon \gamma}{2{b'_1}^2+2b'_1\epsilon \gamma +\gamma^2}\right)x+\left(\frac{2\alpha_0 b'_1+2\alpha_0 \epsilon \gamma +{b'_1}^3\gamma^2}{2{b'_1}^2+2b'_1\epsilon \gamma +\gamma^2}\right)\\&=& \left(x^2+\epsilon \gamma x+\frac{\gamma^2}{2}\right)(x+b'_1-\epsilon \gamma).\end{eqnarray*} 
This shows that $p(x)$ is a non-unit in $\mathcal{R}_{\alpha, \beta}$ if and only if \vspace{-1mm}\begin{equation*} p(x) \in  \left<x^2+ \gamma x+\frac{\gamma^2}{2}\right> \cup \left<x^2- \gamma x+\frac{\gamma^2}{2}\right>,
\text{~~which completes the proof.} 
\vspace{-1mm}\end{equation*} \vspace{-10mm}$\hfill \Box$ \end{description}
As a consequence of the above two propositions, we have the following:
\begin{cor}\label{corr}\begin{enumerate}\item[(a)] There are exactly two maximal ideals of $\mathcal{R}_{\alpha,\beta},$ namely  $\left<x^2+\gamma x +\frac{\gamma^2}{2}\right>$ and $\left<x^2-\gamma x +\frac{\gamma^2}{2}\right>.$\item[(b)] All the nilpotent elements of $\mathcal{R}_{\alpha,\beta}$ are given by $\left<x^4-\alpha_0\right>.$  \item[(b)] All the non-units of $\mathcal{R}_{\alpha,\beta}$ are given by \vspace{-1mm}\begin{equation*}\left<x^2+\gamma x+\frac{\gamma^2}{2}\right>\cup \left<x^2-\gamma x+\frac{\gamma^2}{2}\right>. \vspace{-1mm}\end{equation*}\end{enumerate}
 \end{cor} \noindent \textbf{Proof.}  It follows immediately from Propositions \ref{p1} and \ref{p2}. $\hfill \Box$
 
Working as in Propositions \ref{p1} and \ref{p2}, we also observe the following:
 \begin{prop} \label{p3}\begin{enumerate}\item[(a)] There are exactly two maximal ideals of the quotient ring $\mathbb{F}_{p^m}[x]/\left<x^{4p^s}-\alpha\right>,$ namely $\left<x^2+\gamma x+\frac{\gamma^2}{2}\right>$ and $\left<x^2-\gamma x+\frac{\gamma^2}{2}\right>.$ 
\item[(b)] All the ideals of $\mathbb{F}_{p^m}[x]/\left<x^{4p^s}-\alpha\right>$ are given by \vspace{-1mm}\begin{equation*}\left<\left(x^2+\gamma x+\frac{\gamma^2}{2}\right)^{i}\left(x^2-\gamma x+\frac{\gamma^2}{2}\right)^j\right>,\vspace{-1mm}\end{equation*} where $0 \leq i, j \leq p^s.$ Furthermore, each ideal $\left<\left(x^2+\gamma x+\frac{\gamma^2}{2}\right)^{i}\left(x^2-\gamma x+\frac{\gamma^2}{2}\right)^j\right>$ contains $p^{m(4p^s-2i-2j)}$ elements.
 \item[(c)] All the nilpotent elements of $\mathbb{F}_{p^m}[x]/\left<x^{4p^s}-\alpha\right>$ are given by $\left<x^4-\alpha_0\right>.$ \item[(d)] All the non-units of $\mathbb{F}_{p^m}[x]/\left<x^{4p^s}-\alpha\right>$ are given by \vspace{-1mm}\begin{equation*}\left<x^2+\gamma x+\frac{\gamma^2}{2}\right>\cup \left<x^2-\gamma x+\frac{\gamma^2}{2}\right>.\vspace{-1mm}\end{equation*}\end{enumerate} \end{prop}
\noindent\textbf{Proof.}  Working in a similar manner as in Propositions \ref{p1} and \ref{p2}, the desired result follows. $\hfill \Box$

In the following theorem, we determine all $(\alpha+\beta u)$-constacyclic codes of length $4p^s$ over $R,$ i.e., ideals of the ring $\mathcal{R}_{\alpha,\beta}.$
\begin{theo}\label{t11} Let $p^m \equiv 3~(\text{mod }4)$ and $\beta \neq 0.$ All $(\alpha+\beta u)$-constacyclic codes of length $4p^s$ over $R$ are given by the principal ideals \begin{equation*}\left<\left(x^2+\gamma x+\frac{\gamma^2}{2}\right)^i \left(x^2-\gamma x+\frac{\gamma^2}{2}\right)^{j}\right>,\vspace{-1mm}\end{equation*} where $ 0 \leq i, j \leq 2p^s.$\end{theo}
\noindent \textbf{Proof.} Let $\mathcal{C}$ be an $(\alpha+\beta u)$-constacyclic code of length $4p^s$ over $R,$ i.e.,  an ideal of $\mathcal{R}_{\alpha,\beta}.$ Then  by Lemma \ref{pr6}, we see that $\text{Res}(\mathcal{C}) $ is an $\alpha$-constacyclic code of length $4p^s$ over $\mathbb{F}_{p^m},$ i.e., $\text{Res}(\mathcal{C})$ is an ideal of the ring $\mathbb{F}_{p^m}[x]/\left<x^{4p^s}-\alpha\right>.$  By Proposition \ref{p3}(b), we have  \vspace{-1mm}\begin{equation*}\text{Res}(\mathcal{C})=\left<\left(x^2+\gamma x+\frac{\gamma^2}{2}\right)^{a}\left(x^2-\gamma x+\frac{\gamma^2}{2}\right)^{b}\right>,\vspace{-1mm}\end{equation*} where $0 \leq a, b \leq p^s.$ Further, each codeword $c(x) \in \mathcal{C}$ can be  expressed as \begin{equation*}c(x)=\left(x^2+\gamma x+\frac{\gamma^2}{2}\right)^{a} \left(x^2-\gamma x+\frac{\gamma^2}{2}\right)^{b} A_c(x)+u B_c(x),\end{equation*} where $A_c(x), B_c(x) \in \mathcal{R}_{\alpha,\beta}.$ 
Now as $\left(x^2+\gamma x+\frac{\gamma^2}{2}\right)^{p^s} \left(x^2-\gamma x+\frac{\gamma^2}{2}\right)^{p^s} =\beta u $ in $\mathcal{R}_{\alpha,\beta},$ we can further rewrite \begin{equation*}c(x) =\left(x^2+\gamma x+\frac{\gamma^2}{2}\right)^{a_c} \left(x^2-\gamma x+\frac{\gamma^2}{2}\right)^{b_c}D_c(x),\end{equation*} where $a_c \geq a ,$  $b_c \geq  b $ and $D_c(x) \not \in \left< x^2+\gamma x+\frac{\gamma^2}{2}\right> \cup \left< x^2-\gamma x+\frac{\gamma^2}{2}\right>.$ By Proposition \ref{p2}, we see that $D_c(x)$ is a unit in $\mathcal{R}_{\alpha,\beta}.$  Furthermore, we see, by Proposition \ref{p1}(c), that for $\epsilon \in \{1,-1\},$ the polynomials $x^2+\epsilon\gamma x+\frac{\gamma^2}{2}$ and $x^2-\epsilon\gamma x+\frac{\gamma^2}{2}$ are coprime in $\mathcal{R}_{\alpha,\beta},$ which implies that the polynomials $\big(x^2+\epsilon\gamma x+\frac{\gamma^2}{2}\big)^{\ell}$ and $\big(x^2-\epsilon\gamma x+\frac{\gamma^2}{2}\big)^{2p^s}$ are also coprime in $\mathcal{R}_{\alpha,\beta}$ for each integer $\ell \geq 0.$ Hence there exist $L(x), M(x) \in \mathcal{R}_{\alpha,\beta}$ satisfying $\big(x^2+\epsilon\gamma x+\frac{\gamma^2}{2}\big)^{\ell}L(x)+\big(x^2-\epsilon\gamma x+\frac{\gamma^2}{2}\big)^{2p^s}M(x)=1.$ This, by Proposition \ref{p1}(a) and (c), implies that  
$\big(x^2+\epsilon\gamma x+\frac{\gamma^2}{2}\big)^{\ell+2p^s}L(x)=\left\{1-\big(x^2-\epsilon\gamma x+\frac{\gamma^2}{2}\big)^{2p^s}M(x)\right\} \big(x^2+\epsilon\gamma x+\frac{\gamma^2}{2}\big)^{2p^s}=\big(x^2+\epsilon\gamma x+\frac{\gamma^2}{2}\big)^{2p^s}$ in $\mathcal{R}_{\alpha,\beta}.$ From this, we get $\left< \big(x^2+\epsilon\gamma x+\frac{\gamma^2}{2}\big)^{\ell+2p^s}\right> =\left<\big(x^2+\epsilon\gamma x+\frac{\gamma^2}{2}\big)^{2p^s}\right>$ in $\mathcal{R}_{\alpha,\beta}.$ In view of this, we can assume that both $a_c$ and $b_c $ are less than or equal to $2p^s.$

Now let $i=\max\{a_c: c(x) \in \mathcal{C}\}$ and $j=\max\{b_c: c(x) \in \mathcal{C}\}.$ We note that $0 \leq i , j \leq 2p^s$ and that $\mathcal{C} \subseteq  \left<\left(x^2+\gamma x+\frac{\gamma^2}{2}\right)^{i} \left(x^2-\gamma x+\frac{\gamma^2}{2}\right)^{j}\right>.$ 
Further, there exists $g(x) \in \mathcal{C}$ satisfying $a_g=i$ and $b_g=j.$ That is, we have $g(x) = \left(x^2+\gamma x+\frac{\gamma^2}{2}\right)^{i} \left(x^2-\gamma x+\frac{\gamma^2}{2}\right)^{j}D_g(x),$ where $D_g(x)$ is a unit in $\mathcal{R}_{\alpha,\beta}.$ This implies that $\left(x^2+\gamma x+\frac{\gamma^2}{2}\right)^{i} \left(x^2-\gamma x+\frac{\gamma^2}{2}\right)^{j} = g(x) D_g(x)^{-1}\in \mathcal{C},$ from which it follows that $\mathcal{C}=\left< \left(x^2+\gamma x+\frac{\gamma^2}{2}\right)^{i} \left(x^2-\gamma x+\frac{\gamma^2}{2}\right)^{j}\right>,$ where $0 \leq i,j \leq 2p^s.$
$\hfill \Box$

In the following theorem, we determine cardinalities of all $(\alpha+\beta u)$-constacyclic codes of length $4p^s$ over $R.$
 \begin{theo}\label{card} Let $\mathcal{C}=\left< \left(x^2+\gamma x+\frac{\gamma^2}{2}\right)^{i} \left(x^2-\gamma x+\frac{\gamma^2}{2}\right)^{j}\right>$ be an $(\alpha+\beta u)$-constacyclic code of length $4p^s$ over $R,$ where $0 \leq i,j \leq 2p^s.$ Then we have \begin{equation*}|\mathcal{C}|=p^{m(8p^s-2i-2j)}.\vspace{-1mm}\end{equation*}
\end{theo}
\noindent\textbf{Proof.} By Lemma \ref{pr6}, we see that $|\mathcal{C}|=|\text{Res}(\mathcal{C})||\text{Tor}(\mathcal{C})|.$
\begin{description}\item[(i)] When $i=j=2p^s,$ we see that $\text{Res}(\mathcal{C})=\text{Tor}(\mathcal{C})=\{0\},$ which implies that $|\mathcal{C}|=1.$
\item[(ii)] When $i=j=0,$  we see that  $\text{Res}(\mathcal{C})=\text{Tor}(\mathcal{C})=\mathbb{F}_{p^m}[x]/\left<x^{4p^s}-\alpha\right>.$ By Proposition \ref{p3}(b), we see that the ring $\mathbb{F}_{p^m}[x]/\left<x^{4p^s}-\alpha\right>$ has $p^{4mp^s}$ elements, which gives $|\mathcal{C}|=p^{8mp^s}.$
\item[(iii)] When $p^s \leq i,j \leq 2p^s$ and $(i,j) \neq (2p^s,2p^s),$ we see that \vspace{-1mm}\begin{equation*} \text{Res}(\mathcal{C})=\{0\} \text{ and } \text{Tor}(\mathcal{C})=\left< \left(x^2+\gamma x +\frac{\gamma^2}{2}\right)^{i-p^s} \left(x^2-\gamma x +\frac{\gamma^2}{2}\right)^{j-p^s}\right>.\vspace{-1mm}\end{equation*} By Proposition \ref{p3}(b), we see that $|\text{Res}(\mathcal{C})|=1$ and $|\text{Tor}(\mathcal{C})|=p^{m(8p^s-2i-2j)},$ which gives $|\mathcal{C}|=p^{m(8p^s-2i-2j)}.$
\item[(iv)] When $0 \leq i,j \leq p^s$ and $(i,j) \neq (0,0),$ we see that \vspace{-1mm}\begin{equation*}\text{Res}(\mathcal{C})=\left< \left(x^2+\gamma x +\frac{\gamma^2}{2}\right)^{i} \left(x^2-\gamma x +\frac{\gamma^2}{2}\right)^{j}\right>\text{ and }\text{Tor}(\mathcal{C})=\frac{\mathbb{F}_{p^m}[x]}{\left<x^{4p^s}-\alpha \right>}.\vspace{-1mm}\end{equation*} By Proposition \ref{p3}(b), we see that $|\text{Res}(\mathcal{C})|=p^{m(4p^s-2i-2j)}$ and $|\text{Tor}(\mathcal{C})|=p^{4mp^s},$ which gives $|\mathcal{C}|=p^{m(8p^s-2i-2j)}.$ 
\item[(v)] When $i=j=p^s,$ we see that $\text{Res}(\mathcal{C})=\{0\}$ and $\text{Tor}(\mathcal{C})=\mathbb{F}_{p^m}[x]/\left<x^{4p^s}-\alpha\right>.$ By Proposition \ref{p3}(b), we see that $|\mathcal{C}|=|\text{Tor}(\mathcal{C})|=p^{4mp^s}.$ 
\item[(vi)] Let $0 \leq i \leq p^s$ and $p^s < j \leq 2p^s.$ By Proposition \ref{p1}(c), we see that $\left(x^2+\gamma x +\frac{\gamma^2}{2}\right)^{p^s-i}$ and $ \left(x^2-\gamma x +\frac{\gamma^2}{2}\right)^{j-p^s}$ are coprime in $\mathcal{R}_{\alpha,\beta},$ so there exist $A(x), B(x) \in \mathcal{R}_{\alpha,\beta}$ such that \begin{equation*}\left(x^2+\gamma x +\frac{\gamma^2}{2}\right)^{p^s-i}A(x)+ \left(x^2-\gamma x +\frac{\gamma^2}{2}\right)^{j-p^s} B(x)=1\end{equation*} in $\mathcal{R}_{\alpha,\beta}.$ This implies that \begin{equation*} \left(x^2+\gamma x +\frac{\gamma^2}{2}\right)^{i}\left(x^2-\gamma x +\frac{\gamma^2}{2}\right)^{p^s}+u \beta A(x)=\left(x^2+\gamma x +\frac{\gamma^2}{2}\right)^{i}\left(x^2+\gamma x +\frac{\gamma^2}{2}\right)^{j}B(x) \in \mathcal{C}.\end{equation*} From this, one can show that  \begin{equation*}\text{Res}(\mathcal{C})=\left< \left(x^2+\gamma x +\frac{\gamma^2}{2}\right)^{i} \left(x^2-\gamma x +\frac{\gamma^2}{2}\right)^{p^s}\right>.\end{equation*} Further, as $\left(x^2+\gamma x +\frac{\gamma^2}{2}\right)^{p^s} \left(x^2-\gamma x +\frac{\gamma^2}{2}\right)^{j}= u \beta \left(x^2+\gamma x +\frac{\gamma^2}{2}\right)^{j-p^s} \in \mathcal{C},$ one can observe that \begin{equation*}\text{Tor}(\mathcal{C})= \left<\left(x^2-\gamma x +\frac{\gamma^2}{2}\right)^{j-p^s}\right>.\vspace{-1mm}\end{equation*} Now by Proposition \ref{p3}(b), we note that $|\text{Res}(\mathcal{C})|=p^{m(2p^s-2i)}$ and $|\text{Tor}(\mathcal{C})|=p^{m(6p^s-2j)},$ which gives $|\mathcal{C}|=p^{m(8p^s-2i-2j)}.$ \item[(vii)] When $p^s < i \leq 2p^s$ and $0 \leq j \leq p^s,$ working in a similar manner as in case (vi), we see that \begin{equation*}\text{Res}(\mathcal{C})=\left< \left(x^2+\gamma x +\frac{\gamma^2}{2}\right)^{p^s} \left(x^2-\gamma x +\frac{\gamma^2}{2}\right)^{j}\right>\text{ and } \text{Tor}(\mathcal{C})=\left<\left(x^2+\gamma x +\frac{\gamma^2}{2}\right)^{i-p^s}\right>.\vspace{-1mm}\end{equation*} Now by Proposition \ref{p3}(b), we see that $|\text{Res}(\mathcal{C})|=p^{m(2p^s-2j)}$ and $|\text{Tor}(\mathcal{C})|=p^{m(6p^s-2i)},$ which gives $|\mathcal{C}|=p^{m(8p^s-2i-2j)}.$ \end{description}
$\hfill \Box$

In the following theorem, we determine dual codes of all $(\alpha+\beta u)$-constacyclic codes of length $4p^s$ over $R.$
 \begin{theo} \label{d4} Let $\mathcal{C}=\left< \left(x^2+\gamma x+\frac{\gamma^2}{2}\right)^{i} \left(x^2-\gamma x+\frac{\gamma^2}{2}\right)^{j}\right>$ be an $(\alpha+\beta u)$-constacyclic code of length $4p^s$ over $R,$ where $0 \leq i,j \leq 2p^s.$ We have \begin{equation*}\mathcal{C}^{\perp}=\left<\left(x^2+2\gamma^{-1}x+\frac{2}{\gamma^2}\right)^{2p^s-i}\left(x^2-2\gamma^{-1}x+\frac{2}{\gamma^2}\right)^{2p^s-j}\right>.\vspace{-1mm}\end{equation*}
\end{theo}
\noindent\textbf{Proof.}   To prove this, we first observe that \vspace{-1mm}\begin{equation*}\text{ann}(\mathcal{C})=\left< \left(x^2+\gamma x +\frac{\gamma^2}{2}\right)^{2p^s-i} \left(x^2-\gamma x +\frac{\gamma^2}{2}\right)^{2p^s-j}\right>.\vspace{-1mm}\end{equation*} From this and using Lemma \ref{pr4}, we get \vspace{-1mm}\begin{equation*}\mathcal{C}^{\perp}=\left<\left(x^2+2\gamma^{-1} x +\frac{2}{\gamma^2}\right)^{2p^s-i} \left(x^2-2\gamma^{-1} x +\frac{2}{\gamma^2}\right)^{2p^s-j}\right>.\vspace{-1mm}\end{equation*} $\hfill \Box$\\
In the following corollary, we  determine all isodual $(\alpha+\beta u)$-constacyclic codes of length $4p^s$ over $R.$
\begin{cor} \label{c4} All isodual  $(\alpha+\beta u)$-constacyclic codes of length $4p^s$ over $R$ are given by \begin{equation*}\left< \left(x^2+\gamma x+\frac{\gamma^2}{2}\right)^{i} \left(x^2-\gamma x+\frac{\gamma^2}{2}\right)^{2p^s-i}\right>,\end{equation*} where $0 \leq i \leq 2p^s.$\end{cor}
\noindent \textbf{Proof.} To prove this,  let $\mathcal{C}$ be an $(\alpha+\beta u)$-constacyclic code of length $4p^s$ over $R.$ By Theorem \ref{t11}, we see that $\mathcal{C}= \left< \left(x^2+\gamma x+\frac{\gamma^2}{2}\right)^{i} \left(x^2-\gamma x+\frac{\gamma^2}{2}\right)^{j}\right>,$ where $0 \leq i,j \leq 2p^s.$ Further, by Theorems \ref{card} and \ref{d4}, we get $|\mathcal{C}|=p^{m(8p^s-2i-2j)},$ $\mathcal{C}^{\perp}=\left<\left(x^2+2\gamma^{-1}x+\frac{2}{\gamma^2}\right)^{2p^s-i}\left(x^2-2\gamma^{-1}x+\frac{2}{\gamma^2}\right)^{2p^s-j}\right>$ and $|\mathcal{C}^{\perp}|=p^{m(2i+2j)}.$ Now for the code $\mathcal{C}$ to be isodual, we must have $|\mathcal{C}|=|\mathcal{C}^{\perp}|,$ which gives $i+j=2p^s.$ On the other hand, it is easy to see that for $0 \leq i \leq 2p^s,$ the code $\left< \left(x^2+\gamma x+\frac{\gamma^2}{2}\right)^{i} \left(x^2-\gamma x+\frac{\gamma^2}{2}\right)^{2p^s-i}\right>$ is isodual. From this, the desired result follows immediately. $\hfill \Box$
\subsection{The case $\beta =0$}
In this section, we shall determine all $\alpha$-constacyclic codes of length $4p^s$ over $R,$ where  $\alpha \in \mathbb{F}_{p^m}\setminus \{0\}$ is not a square in $\mathbb{F}_{p^m}.$ For this, we see that all $\alpha$-constacyclic codes of length $4p^s$ over $R$ are ideals of the quotient ring $\mathfrak{R}_{\alpha}=R[x]/\left<x^{4p^s}-\alpha\right>.$ By Lemma \ref{pr5}(c), there exists $\alpha_0 \in \mathbb{F}_{p^m}$ such that $\alpha_0^{p^s}=\alpha.$ Further, by Proposition \ref{p1}(b), the equation $x^4+ 4 \alpha_0=0$ has exactly two roots in $\mathbb{F}_{p^m}.$   In fact, if $\gamma$ is a root of $x^4+4 \alpha_0=0$ in $\mathbb{F}_{p^m},$ then $-\gamma$ is its other root. We further observe that $x^4-\alpha_0=\left(x^2+\gamma x+\frac{\gamma^2}{2}\right) \left(x^2-\gamma x+\frac{\gamma^2}{2}\right),$ which implies that $x^{4p^s}-\alpha=\left(x^2+\gamma x+\frac{\gamma^2}{2}\right)^{p^s} \left(x^2-\gamma x+\frac{\gamma^2}{2}\right)^{p^s}.$ Furthermore,  one can easily observe that the ideals \small{$\left<\left(x^2+\gamma x+\frac{\gamma^2}{2}\right)^{p^s}\right>$~}\normalsize and \small{$\left<\left(x^2-\gamma x+\frac{\gamma^2}{2}\right)^{p^s}\right>$ }\normalsize are  coprime in $R[x].$ So by Chinese Remainder Theorem, we have \small{\vspace{-1mm}\begin{equation*}\mathfrak{R}_{\alpha}=\frac{R[x]}{\left<\left(x^2+\gamma x+\frac{\gamma^2}{2}\right)^{p^s}\right>} \oplus \frac{R[x]}{\left< \left(x^2-\gamma x+\frac{\gamma^2}{2}\right)^{p^s}\right>}=\mathsf{R}_{\gamma}\oplus \mathsf{R}_{-\gamma},\vspace{-1mm}\end{equation*}}\normalsize where $\mathsf{R}_{\gamma}=\frac{\displaystyle R[x]}{\left<\left(x^2+\gamma x+\frac{\gamma^2}{2}\right)^{p^s}\right>}~$ and  $~ \mathsf{R}_{-\gamma}=\frac{\displaystyle R[x]}{\left< \left(x^2-\gamma x+\frac{\gamma^2}{2}\right)^{p^s}\right>}.$ Now the following result is well-known:
\begin{lem}\label{l7}\begin{enumerate}\item[(a)] Let $I$ be an $\alpha$-constacyclic code of length $4p^s$ over $R,$ i.e.,  an ideal of the ring $\mathfrak{R}_{\alpha}.$ Then $I=I_1\oplus I_2,$ where $I_1$ is an ideal of $\mathsf{R}_{\gamma}$ and $I_2$ is an ideal of $\mathsf{R}_{-\gamma}.$
\item[(b)] If $I_1$ is an ideal of $\mathsf{R}_{\gamma}$ and $I_2$ is an ideal of $\mathsf{R}_{-\gamma},$ then $I=I_1 \oplus I_2$ is an ideal of $\mathfrak{R}_{\alpha},$ (i.e., $I$ is an $\alpha$-constacyclic code of length $4p^s$ over $R$).  Moreover, we have $|I|=|I_1||I_2|.$\end{enumerate}\end{lem}\noindent\textbf{Proof.} Proof is trivial. $\hfill\Box$

In view of the above lemma, we see that  to determine all $\alpha$-constacyclic codes of length $4p^s$ over $R,$ it is enough to determine all ideals of the rings $\mathsf{R}_{\gamma}$ and $\mathsf{R}_{-\gamma}.$
For this, we first prove the following lemmas:
\begin{lem}\label{l9} Let $\epsilon \in \mathbb{F}_{p^m}$ be either 1 or $-1.$ Then the following hold.
\begin{enumerate}\item[(a)] Any non-zero polynomial $p(x)=ax+b$ is a unit in $\mathsf{R}_{\epsilon \gamma},$ where $a,b \in \mathbb{F}_{p^m}.$
\item[(b)]  In $\mathsf{R}_{\epsilon \gamma},$ the element $x^2+\epsilon \gamma x+\frac{\gamma^2}{2}$ is nilpotent and its nilpotency index is $p^s.$  \item[(c)] The ring $\mathsf{R}_{\epsilon \gamma}$ is a finite commutative local ring with the maximal ideal as $\left<x^2+\epsilon \gamma x+\frac{\gamma^2}{2},u\right>.$ As a consequence,  $\mathsf{R}_{\epsilon \gamma}$ is not a chain ring.\end{enumerate}\end{lem}
\noindent \textbf{Proof.} (a) When $a=0,$ we have $p(x)=b \in \mathbb{F}_{p^m}\setminus \{0\},$ which is trivially a unit in $\mathsf{R}_{\epsilon \gamma}.$ So we assume that $a$ is non-zero and  we write $ p(x)=(ax+b)=a(x+a^{-1}b)=a(x+b_1),$ where $b_1=a^{-1}b.$ Here we see that $b_1^2-\epsilon b_1\gamma +\frac{\gamma^2}{2}$ is a non-zero element of $\mathbb{F}_{p^m}$ for any $b_1 \in \mathbb{F}_{p^m}.$ Further, we observe that \vspace{-1mm}\begin{equation*}-p(x) a^{-1}(x+b_1)^{p^s-1}(x+\epsilon \gamma -b_1)^{p^s}\left\{b_1^2-\epsilon b_1\gamma +\frac{\gamma^2}{2}\right\}^{-p^s}=1\text{ in }\mathsf{R}_{\epsilon \gamma}.\vspace{-1mm}\end{equation*}  This shows that $p(x)$ is a unit in $\mathsf{R}_{\epsilon \gamma}.$\\(b) It follows from the fact that $\left(x^2+\epsilon\gamma x+\frac{\gamma^2}{2}\right)^{p^s}=0$ in $\mathsf{R}_{\epsilon\gamma}.$\\(c) To prove this, we see that any element $f(x) \in \mathsf{R}_{\epsilon \gamma}$ can be uniquely expressed as \vspace{-1mm}\begin{equation*}f(x)=\sum\limits_{i=0}^{p^s-1}(a_{i0}x+b_{i0})(x^2+\epsilon \gamma x+\frac{\gamma^2}{2})^i+ u \sum\limits_{i=0}^{p^s-1}(a_{i1}x+b_{i1})(x^2+\epsilon \gamma x+\frac{\gamma^2}{2})^i,\vspace{-1mm}\end{equation*} where $a_{ij},b_{ij} \in \mathbb{F}_{p^m}$ for each $i$ and $j.$ Now by parts (a) and (b), we see that $f(x)$ is a unit in $\mathsf{R}_{\epsilon \gamma}$ if and only if  $a_{00}x+b_{00}$ is non-zero. From this, it follows that  the ideal $\left<x^2+\epsilon \gamma x+\frac{\gamma^2}{2},u\right>$ consists of all the non-units in $\mathsf{R}_{\epsilon \gamma}.$  This implies that $\mathsf{R}_{\epsilon \gamma}$  is a local ring with the maximal ideal as $\left<x^2+\epsilon \gamma x+\frac{\gamma^2}{2},u\right>,$ which is not a principal ideal. Therefore by Proposition \ref{pr1}, $\mathsf{R}_{\epsilon \gamma}$  is not a chain ring. $\hfill \Box$
\begin{lem} \label{l12}Let $\epsilon \in \mathbb{F}_{p^m}$ be either 1 or $-1.$  The quotient ring $ \mathbb{F}_{p^m}[x]/\left<(x^2+\epsilon \gamma x+\frac{\gamma^2}{2})^{p^s}\right>$ is a finite commutative local ring with unity and having the maximal ideal as $\left<x^2+\epsilon \gamma x+\frac{\gamma^2}{2}\right>.$ As a consequence,  $ \mathbb{F}_{p^m}[x]/\left<(x^2+\epsilon \gamma x+\frac{\gamma^2}{2})^{p^s}\right>$ is a chain ring and all its ideals are given by $\left<(x^2+\epsilon \gamma x+\frac{\gamma^2}{2})^i\right>,$ where $0 \leq i \leq p^s.$ Furthermore, $|\left<(x^2+\epsilon \gamma x+\frac{\gamma^2}{2})^i\right> |=p^{2m(p^s-i)}$ for $0 \leq i \leq p^s.$\end{lem}
\noindent\textbf{Proof.} To prove this, we see that any element $f(x) \in  \mathbb{F}_{p^m}[x]/\left<(x^2+\epsilon \gamma x+\frac{\gamma^2}{2})^{p^s}\right>$ can be uniquely expressed as $f(x)=\sum\limits_{j=0}^{p^s-1}(a_{j0}x+b_{j0})(x^2+\epsilon \gamma x+\frac{\gamma^2}{2})^j,$ where $a_{j0},b_{j0} \in \mathbb{F}_{p^m}$ for $0 \leq j \leq p^s-1.$ As $x^2+\epsilon \gamma x+\frac{\gamma^2}{2}$ is a nilpotent element of $\mathbb{F}_{p^m}[x]/\left<(x^2+\epsilon \gamma x+\frac{\gamma^2}{2})^{p^s}\right>,$ we observe that $f(x)$ is a unit in $\mathbb{F}_{p^m}[x]/\left<(x^2+\epsilon \gamma x+\frac{\gamma^2}{2})^{p^s}\right>$ if and only if $a_{00}x+b_{00}$ is a unit in $\mathbb{F}_{p^m}[x]/\left<(x^2+\epsilon \gamma x+\frac{\gamma^2}{2})^{p^s}\right>.$ Working in a similar manner as in Lemma \ref{l9}, we see that $a_{00}x+b_{00}$ is  a unit in $\mathbb{F}_{p^m}[x]/\left<(x^2+\epsilon \gamma x+\frac{\gamma^2}{2})^{p^s}\right>$ if and only if $a_{00}x+b_{00}$ is non-zero. From this, it follows that  the ideal $\left<x^2+\epsilon \gamma x+\frac{\gamma^2}{2}\right>$ consists of all the non-units in $\mathbb{F}_{p^m}[x]/\left<(x^2+\epsilon \gamma x+\frac{\gamma^2}{2})^{p^s}\right>.$ Therefore $\mathbb{F}_{p^m}[x]/\left<(x^2+\epsilon \gamma x+\frac{\gamma^2}{2})^{p^s}\right>$ is a local ring with the maximal ideal as $\left<x^2+\epsilon \gamma x+\frac{\gamma^2}{2}\right>.$ Now using Proposition \ref{pr1}, the desired result follows. $\hfill \Box$
 \begin{theo}\label{t7} Let $\epsilon \in \mathbb{F}_{p^m}$ be either 1 or $-1.$ All the  ideals of $\mathsf{R}_{\epsilon \gamma}$ are as listed below:
\begin{enumerate}\addtolength{\itemindent}{1.0cm}\item[Type I:] (Trivial ideals)
\vspace{-1mm}\begin{equation*}\{0\},~~~\mathsf{R}_{\epsilon \gamma}. \vspace{-1mm}\end{equation*}
\item[Type II:] (Principal ideals with non-monic polynomial generators)
\vspace{-1mm}\begin{equation*} \left<u(x^2+\epsilon \gamma x+\frac{\gamma^2}{2})^i \right>, ~~\text{where~~} 0 \leq i \leq p^s-1.\vspace{-1mm}\end{equation*}
\item[Type III:] (Principal ideals with monic polynomial generators)
\vspace{-1mm}\begin{equation*}\left<(x^2+\epsilon \gamma x+\frac{\gamma^2}{2})^i+u(x^2+\epsilon \gamma x+\frac{\gamma^2}{2})^t h(x)\right>,\vspace{-1mm}\end{equation*} where $1 \leq i \leq p^s-1, ~0 \leq t <i$ and either $h(x)$ is 0 or $h(x)$ is a unit in $\mathsf{R}_{\epsilon \gamma}$ that can be represented as $h(x)=\sum\limits_{j=0}^{i-t-1}(a_{j0}x+b_{j0})(x^2+\epsilon \gamma x+\frac{\gamma^2}{2})^j$ with each $a_{j0},b_{j0}\in \mathbb{F}_{p^m}$ and $a_{00}x+b_{00} \neq 0.$
\item[Type IV:] (Non-principal ideals)
\vspace{-1mm}\begin{equation*}\left<(x^2+\epsilon \gamma x+\frac{\gamma^2}{2})^i+u(x^2+\epsilon \gamma x+\frac{\gamma^2}{2})^t h(x), u(x^2+\epsilon \gamma x+\frac{\gamma^2}{2})^w\right>,\vspace{-1mm}\end{equation*}
where $1 \leq i \leq p^s-1, ~0 \leq t <w,$ either $h(x)$ is 0 or $h(x)$ is a unit in $\mathsf{R}_{\epsilon \gamma}$ that can be represented as $h(x)=\sum\limits_{j=0}^{w-t-1}(a_{j0}x+b_{j0})(x^2+\epsilon \gamma x+\frac{\gamma^2}{2})^j$ with each $a_{j0},b_{j0} \in \mathbb{F}_{p^m},$ $a_{00}x+b_{00} \neq 0$ and $w < U$ with $U$ as the smallest integer such that $u(x^2+\epsilon \gamma x+\frac{\gamma^2}{2})^U \in \left<(x^2+\epsilon \gamma x+\frac{\gamma^2}{2})^i+u(x^2+\epsilon \gamma x+\frac{\gamma^2}{2})^t h(x)\right>.$ \end{enumerate}
\end{theo}
\noindent\textbf{Proof.} Working in a similar manner as in Theorem 3.9 of Chen et al. \cite{chen}, the desired result follows. $\hfill \Box$
\\In the following proposition, we determine the integer $U$ as defined in Theorem \ref{t7}.
\begin{prop}\label{U} Let $\epsilon \in \mathbb{F}_{p^m}$ be either 1 or $-1.$ The smallest integer $U$ satisfying $u(x^2+\epsilon \gamma x+\frac{\gamma^2}{2})^U \in \left<(x^2+\epsilon \gamma x+\frac{\gamma^2}{2})^i+u(x^2+\epsilon \gamma x+\frac{\gamma^2}{2})^t h(x)\right>$ is given by \vspace{-1mm}\begin{equation*}U=\left\{\begin{array}{cl} i & \text{if }h(x)=0;\\
\min\{i,p^s-i+t\} & \text{if }h(x) \text{ is a unit in }\mathsf{R}_{\epsilon \gamma}. \end{array}\right.\vspace{-1mm}\end{equation*} \end{prop}
\noindent\textbf{Proof.} Working in a similar manner as in Proposition 3.10 of Chen et al. \cite{chen}, the result follows. $\hfill \Box$

In the following theorem, we  determine cardinalities of all ideals of the ring $\mathsf{R}_{\epsilon \gamma},$ where $\epsilon \in \mathbb{F}_{p^m}$ is either 1 or $-1.$
\begin{theo}\label{t5} Let $\epsilon \in \mathbb{F}_{p^m}$ be either 1 or $-1.$ Let $\mathcal{I}$ be an ideal of the ring $\mathsf{R}_{\epsilon \gamma}.$  Then following the same notations as in Theorem \ref{t7}, we have the following:
\begin{enumerate}
\item[(a)] If $\mathcal{I}=\{0\},$ then $|\mathcal{I}|=1.$
\item[(b)] If $\mathcal{I}=\mathsf{R}_{\epsilon\gamma},$ then $|\mathcal{I}|=p^{4 mp^s}.$
\item[(c)] If $\mathcal{I}=\left<u(x^2+\epsilon \gamma x+\frac{\gamma^2}{2})^i \right>$ with $0 \leq i \leq p^s-1,$ then $|\mathcal{I}|=p^{2m(p^s-i)}.$
\item[(d)] If $\mathcal{I}=\left<(x^2+\epsilon \gamma x+\frac{\gamma^2}{2})^i+u(x^2+\epsilon \gamma x+\frac{\gamma^2}{2})^t h(x)\right>$ is of the Type III,  then  \vspace{-1mm}\begin{equation*}|\mathcal{I}|=\left\{\begin{array}{ll}p^{4m(p^s-i)} & \text{if either }h(x) =0 \text{ or }h(x)\text{ is a unit in }\mathsf{R}_{\epsilon\gamma} \text{ and }1 \leq i \leq \frac{p^s+t}{2};\\
p^{2m(p^s-t)} & \text{if }h(x)\text{ is a unit in }\mathsf{R}_{\epsilon\gamma} \text{ and }\frac{p^s+t}{2}< i \leq p^s-1.\end{array}\right.\vspace{-1mm}\end{equation*}
\item[(e)]  If $\mathcal{I}=\left<(x^2+\epsilon \gamma x+\frac{\gamma^2}{2})^i+u(x^2+\epsilon \gamma x+\frac{\gamma^2}{2})^t h(x), u(x^2+\epsilon \gamma x+\frac{\gamma^2}{2})^w\right>$
is of the Type IV,  then  \vspace{-1mm}\begin{equation*} |\mathcal{I}|=p^{2m(2p^s-i-w)}. \vspace{-1mm}\end{equation*}
 \end{enumerate}\end{theo}
 \noindent \textbf{Proof.}  To prove this, working as in Lemma \ref{pr6}, we observe that $|\mathcal{I}|=|\text{Res}(\mathcal{I})||\text{Tor}(\mathcal{I})|.$
\\(a) When $\mathcal{I}=\{0\},$ then $\text{Res}(\mathcal{I})=\text{Tor}(\mathcal{I})=\{0\},$ which implies that $|\mathcal{I}|=1.$
\\(b) When $\mathcal{I}=\mathsf{R}_{\epsilon \gamma},$ we have $\text{Res}(\mathcal{I})=\text{Tor}(\mathcal{I})=\mathbb{F}_{p^m}[x]/\left<\left(x^2+\epsilon \gamma x+\frac{\gamma^2}{2}\right)^{p^s}\right>,$ which, by Lemma \ref{l12}, implies that $|\mathcal{I}|=p^{4mp^s}.$
\\(c) When $\mathcal{I}=\left<u(x^2+\epsilon \gamma x+\frac{\gamma^2}{2})^i \right>,$ we have $\text{Res}(\mathcal{I})=\{0\}$ and $\text{Tor}(\mathcal{I})=\left<(x^2+\epsilon \gamma x+\frac{\gamma^2}{2})^i \right>.$ This, by Lemma \ref{l12}, implies that $|\mathcal{I}|=p^{2m(p^s-i)}.$\\(d) When $\mathcal{I}=\left<(x^2+\epsilon \gamma x+\frac{\gamma^2}{2})^i+u(x^2+\epsilon \gamma x+\frac{\gamma^2}{2})^t h(x)\right>,$ we have $\text{Res}(\mathcal{I})=\left<(x^2+\epsilon \gamma x+\frac{\gamma^2}{2})^i\right>$ and $$\text{Tor}(\mathcal{I})=\left\{\begin{array}{ll} \left<(x^2+\epsilon \gamma x+\frac{\gamma^2}{2})^i \right> & \text{if either } h(x)=0 \text{ or } h(x) \text{ is a unit in } \mathsf{R}_{\epsilon\gamma} \text{ and } 1 \leq i \leq \frac{p^s+t}{2};\\ \left<  (x^2+\epsilon \gamma x+\frac{\gamma^2}{2})^{p^s-i+t}\right> & \text{if }h(x) \text{ is a unit in } \mathsf{R}_{\epsilon\gamma} \text{ and } \frac{p^s+t}{2} < i \leq p^s-1.\end{array}\right.$$ From this and by applying Lemma \ref{l12}, part (d) follows.\\(e)  When $\mathcal{I}=\left<(x^2+\epsilon \gamma x+\frac{\gamma^2}{2})^i+u(x^2+\epsilon \gamma x+\frac{\gamma^2}{2})^t h(x), u(x^2+\epsilon \gamma x+\frac{\gamma^2}{2})^w\right>,$ we have $$\text{Res}(\mathcal{I})=\left<(x^2+\epsilon \gamma x+\frac{\gamma^2}{2})^i\right>\text{ and }\text{Tor}(\mathcal{I})=\left<(x^2+\epsilon \gamma x+\frac{\gamma^2}{2})^w \right>.$$ From this and by applying by Lemma \ref{l12}, part (e) follows. $\hfill \Box$\\
Next we see that if   $I \subseteq \mathfrak{R}_{\alpha}$ is an $\alpha$-constacyclic code of length $4p^s$ over $R,$ then its dual code $I^{\perp}$ is an ideal of the ring $\mathfrak{R}_{\alpha^{-1}}.$ As $\alpha=\alpha_0^{p^s}$ and $-4\alpha_0=\gamma^4,$ we can write $x^{4p^s}-\alpha^{-1}=\left(x^4+\frac{4}{\gamma^4}\right)^{p^s}=\left(x^2+2\gamma^{-1}x+\frac{2}{\gamma^2}\right)^{p^s}\left(x^2-2\gamma^{-1}x+\frac{2}{\gamma^2}\right)^{p^s}.$ Further, we see that the ideals $\left< \left(x^2+2\gamma^{-1}x+\frac{2}{\gamma^2}\right)^{p^s}\right>$ and $\left<\left(x^2-2\gamma^{-1}x+\frac{2}{\gamma^2}\right)^{p^s} \right>$ are coprime in $\mathfrak{R}_{\alpha^{-1}}.$ Therefore  by Chinese Remainder Theorem, we have \small{\vspace{-1mm}\begin{equation*}\mathfrak{R}_{\alpha^{-1}}=\frac{R[x]}{\left<\left(x^2+2\gamma^{-1}x+\frac{2}{\gamma^2}\right)^{p^s}\right>} \oplus \frac{R[x]}{\left<\left(x^2-2\gamma^{-1}x+\frac{2}{\gamma^2}\right)^{p^s}\right>}=\mathsf{R}_{2\gamma^{-1}}\oplus \mathsf{R}_{-2\gamma^{-1}}.\vspace{-1mm}\end{equation*}}\normalsize In the following lemma, we relate dual codes of $\alpha$-constacyclic codes of length $4p^s$ over $R$ with the orthogonal complements of ideals of the rings $\mathsf{R}_{\gamma}$ and $\mathsf{R}_{-\gamma}.$
\begin{lem}\label{d3} Let $I$ be an $\alpha$-constacyclic code of length $4p^s$ over $R,$ i.e.,  an ideal of the ring $\mathfrak{R}_{\alpha}.$ If $I=I_1\oplus I_2$ with $I_1$ an ideal of $\mathsf{R}_{\gamma}$ and $I_2$ an ideal of $\mathsf{R}_{-\gamma},$ then the dual code $I^{\perp}$ of $I$ is given by $I^{\perp}=I_1^{\perp}\oplus I_2^{\perp},$ where $I_1^{\perp} $ is the orthogonal complement of  $I_1$  and  $I_2^{\perp}$ is the orthogonal complement of $I_2.$ Furthermore, $I_1^{\perp}$ is an ideal of $\mathsf{R}_{2\gamma^{-1}}$ and $I_2^{\perp}$ is an ideal of $\mathsf{R}_{-2 \gamma^{-1}}.$  \end{lem}
\noindent\textbf{Proof.}   Its proof is straightforward. $\hfill \Box$\\
In view of the above lemma, we see that to determine dual codes of all $\alpha$-constacyclic codes of length $4p^s$ over $R,$ it is enough to determine orthogonal complements of all ideals of the rings $\mathsf{R}_{\gamma}$ and $\mathsf{R}_{-\gamma}.$ Towards this, we make the following observation:
\begin{lem} \label{l13} Let $\epsilon \in \mathbb{F}_{p^m}$ be either 1 or $-1.$ Let $\mathcal{I}$ be an ideal of the ring $\mathsf{R}_{\epsilon \gamma}$ with the orthogonal complement as $\mathcal{I}^{\perp}.$ Then the following hold.
\begin{enumerate}\item[(a)] $\mathcal{I}^{\perp}$ is an ideal of $\mathsf{R}_{2\epsilon \gamma^{-1}}.$  \item[(b)] $\mathcal{I}^{\perp}=\text{ann}(\mathcal{I})^*.$\item[(c)] If $\mathcal{I}=\left<f(x),ug(x)\right>,$ then $\mathcal{I}^{\perp}=\left<f^*(x),ug^*(x)\right>.$
\item[(d)] For $f(x),g(x) \in \mathsf{R}_{\epsilon \gamma},$ we have $\left<\left(f(x)g(x)\right)^*\right>=\left<f^*(x)g^*(x)\right>.$
\item[(e)] For non-zero $f(x),g(x) \in \mathsf{R}_{\epsilon \gamma}$ with $f(x)+g(x) \neq 0$ and $\text{deg }f(x) \geq \text{deg }g(x),$ we have \vspace{-1mm}\begin{equation*}\left<\left(f(x)+g(x)\right)^*\right>=\left<f^*(x)+x^{\text{deg }f(x)-\text{deg }g(x)}g^*(x)\right>.\vspace{-1mm}\end{equation*} \end{enumerate}\end{lem}
\noindent \textbf{Proof.} Its proof is similar to that of Lemmas \ref{pr2} and  \ref{pr4}. $\hfill \Box$
\\In the following theorem, we determine orthogonal complements of all ideals of the ring $\mathsf{R}_{\epsilon \gamma},$ where $\epsilon \in \mathbb{F}_{p^m}$ is either $1$ or $-1.$
\begin{theo} \label{tt6} Let $\epsilon \in \mathbb{F}_{p^m}$ be either 1 or $-1.$ Let $\mathcal{I}$ be an ideal of the ring $\mathsf{R}_{\epsilon \gamma}.$ Then following the same notations as in Theorem \ref{t7}, we have the following:
\begin{enumerate}
\item[(a)] If $\mathcal{I}=\{0\},$ then $\mathcal{I}^{\perp}=\mathsf{R}_{2\epsilon \gamma^{-1}}$
\item[(b)] If $\mathcal{I}=\mathsf{R}_{\epsilon\gamma},$ then $\mathcal{I}^{\perp}=\{0\}.$
\item[(c)] If $\mathcal{I}=\left<u(x^2+\epsilon \gamma x+\frac{\gamma^2}{2})^i \right>$ with $0 \leq i \leq p^s-1,$ then $\mathcal{I}^{\perp}=\left<(x^2+2\epsilon \gamma^{-1} x+\frac{2}{\gamma^{2}})^{p^s-i},u \right>.$
\item[(d)] If $\mathcal{I}=\left<(x^2+\epsilon \gamma x+\frac{\gamma^2}{2})^i+u(x^2+\epsilon \gamma x+\frac{\gamma^2}{2})^t h(x)\right>$ is of the Type III,  then  
\vspace{-1mm}\begin{equation*}\mathcal{I}^{\perp}=\left\{\begin{array}{ll}
\left<(x^2+2\epsilon \gamma^{-1} x+\frac{2}{\gamma^{2}})^{p^s-i} \right> \text{if~ }h(x)=0;\\  
\left< (x^2+2\epsilon \gamma^{-1} x+\frac{2}{\gamma^{2}})^{p^s-i}-u x^{2i-2t-\text{deg }h(x)}(\frac{2}{\gamma^2})^{i-t} (x^2+2\epsilon \gamma^{-1} x+\frac{2}{\gamma^{2}})^{p^s-2i+t}h^*(x)\right> \text{ if } h(x)\\ \text{ is a unit in } \mathsf{R}_{\epsilon\gamma}\text{ and }1 \leq i \leq \frac{p^s+t}{2};\\
\left<(\frac{\gamma^2}{2})^{i-t}(x^2+2\epsilon \gamma^{-1} x+\frac{2}{\gamma^{2}})^{i-t}-u x^{2i-2t-\text{deg }h(x)} h^*(x) ,  u (x^2+2\epsilon \gamma^{-1} x+\frac{2}{\gamma^{2}})^{p^s-i}\right>\text{ if } h(x)\\ \text{ is a unit in } \mathsf{R}_{\epsilon\gamma} \text{ and } \frac{p^s+t}{2}< i \leq p^s-1.
\end{array}\right. \vspace{-1mm}\end{equation*}
\item[(e)]  If $\mathcal{I}=\left<(x^2+\epsilon \gamma x+\frac{\gamma^2}{2})^i+u(x^2+\epsilon \gamma x+\frac{\gamma^2}{2})^t h(x), u(x^2+\epsilon \gamma x+\frac{\gamma^2}{2})^w\right>$ is of the Type IV,   then
\vspace{-1mm}\begin{equation*}\mathcal{I}^{\perp}=\left\{\begin{array}{ll}
\left<(x^2+2\epsilon \gamma^{-1} x+\frac{2}{\gamma^{2}})^{p^s-w}, u(x^2+2\epsilon \gamma^{-1} x+\frac{2}{\gamma^{2}})^{p^s-i} \right> \text{if~ }h(x)=0;\\  
\Big<(x^2+2\epsilon \gamma^{-1} x+\frac{2}{\gamma^{2}})^{p^s-w}-ux^{2i-2t-\text{deg }h(x)}  (\frac{2}{\gamma^2})^{i-t} (x^2+2\epsilon \gamma^{-1} x+\frac{2}{\gamma^{2}})^{p^s-i-w+t}h^*(x), \\ u(x^2+2\epsilon \gamma^{-1} x+\frac{2}{\gamma^{2}})^{p^s-i}\Big> \text{ if } h(x) \text{ is a unit in } \mathsf{R}_{\epsilon\gamma}.
\end{array}\right. \vspace{-1mm}\end{equation*}
\end{enumerate}\end{theo}
 \noindent \textbf{Proof.} 
 Proofs of parts (a) and (b) are trivial. \vspace{2mm}\\(c) Let $\mathcal{I}=\left<u(x^2+\epsilon \gamma x+\frac{\gamma^2}{2})^i \right>$ for some $i,~0 \leq i \leq p^s-1.$ Then it is easy to see that \vspace{-1mm}\begin{equation*}\text{ann}(\mathcal{I})=\left<(x^2+\epsilon \gamma x+\frac{\gamma^2}{2})^{p^s-i},u\right>.\vspace{-1mm}\end{equation*} Now using Lemma \ref{l13}, we get \vspace{-1mm}\begin{equation*}\mathcal{I}^{\perp}=\text{ann}(\mathcal{I})^*=\left<(x^2+2\epsilon \gamma^{-1} x+\frac{2}{\gamma^{2}})^{p^s-i},u \right>.\vspace{-1mm}\end{equation*}
\vspace{2mm}\\(d) Let $\mathcal{I}=\left<(x^2+\epsilon \gamma x+\frac{\gamma^2}{2})^i+u(x^2+\epsilon \gamma x+\frac{\gamma^2}{2})^t h(x)\right>,$ where $1 \leq i \leq p^s-1, ~0 \leq t <i$ and $h(x)=\sum\limits_{j=0}^{i-t-1}A_j(x)(x^2+\epsilon \gamma x+\frac{\gamma^2}{2})^j$ (each $A_j(x) \in \mathbb{F}_{p^m}[x]$ is either 0 or $\text{deg }A_j(x) \leq 1$) as either 0 or  a unit in $\mathsf{R}_{\epsilon\gamma}.$

 First let $h(x)=0.$ Then we have $\mathcal{I}=\left<(x^2+\epsilon \gamma x+\frac{\gamma^2}{2})^i\right>.$ Here we observe that \vspace{-1mm}\begin{equation*}\text{ann}(\mathcal{I})=\left<(x^2+\epsilon \gamma x+\frac{\gamma^2}{2})^{p^s-i}\right>,\vspace{-1mm}\end{equation*} which, by Lemma \ref{l13}, implies that \vspace{-1mm}\begin{equation*}\mathcal{I}^{\perp}=\text{ann}(\mathcal{I})^*=\left<(x^2+2\epsilon \gamma^{-1} x+\frac{2}{\gamma^{2}})^{p^s-i}\right>.\vspace{-1mm}\end{equation*}
Next suppose that $h(x)$ is a unit in $\mathsf{R}_{\epsilon\gamma}.$ In order to determine $\mathcal{I}^{\perp},$ we see that  $\text{ann}(\mathcal{I})$ is an ideal of $\mathsf{R}_{\epsilon \gamma}.$ So by Theorem \ref{t7}, we can write 
 \vspace{-1mm}\begin{equation*}\text{ann}(\mathcal{I})=\left<(x^2+\epsilon \gamma x+\frac{\gamma^2}{2})^a+u(x^2+\epsilon \gamma x+\frac{\gamma^2}{2})^b g(x), u (x^2+\epsilon \gamma x+\frac{\gamma^2}{2})^c\right>,\vspace{-1mm}\end{equation*} where $g(x)$ is either 0 or a unit in $\mathsf{R}_{\epsilon \gamma},$ $c$ is the smallest integer satisfying \vspace{-1mm}\begin{equation*} u(x^2+\epsilon \gamma x+\frac{\gamma^2}{2})^{c}\in \Big< (x^2+\epsilon \gamma x+\frac{\gamma^2}{2})^a+u(x^2+\epsilon \gamma x+\frac{\gamma^2}{2})^b g(x)\Big>\vspace{-1mm}\end{equation*} and $a,b$ are integers satisfying $1 \leq a \leq p^s-1$ and $0 \leq b < c.$  This implies that \small {\begin{equation}\label{11} u(x^2+\epsilon \gamma x+\frac{\gamma^2}{2})^{i+c}=0 \text{ and }(x^2+\epsilon \gamma x+\frac{\gamma^2}{2})^{a+i}+u\left\{(x^2+\epsilon \gamma x+\frac{\gamma^2}{2})^{a+t}h(x)+(x^2+\epsilon \gamma x+\frac{\gamma^2}{2})^{i+b}g(x)\right\}=0.\end{equation}}\normalsize
First suppose that $1 \leq i \leq \frac{p^s+t}{2}.$ In this case, we see that \eqref{11} holds for $a=c=p^s-i,$ $g(x)=-h(x)$ and $b=p^s+t-2i.$  This implies that \vspace{-1mm}\begin{equation*}\text{ann}(\mathcal{I})=\Big<(x^2+\epsilon \gamma x+\frac{\gamma^2}{2})^{p^s-i}-u(x^2+\epsilon \gamma x+\frac{\gamma^2}{2})^{p^s+t-2i} h(x), u (x^2+\epsilon \gamma x+\frac{\gamma^2}{2})^{p^s-i}\Big>.\vspace{-1mm}\end{equation*} We note that 
 \vspace{-1mm}\begin{equation*} u(x^2+\epsilon \gamma x+\frac{\gamma^2}{2})^{p^s-i}=u\Big\{(x^2+\epsilon \gamma x+\frac{\gamma^2}{2})^{p^s-i}- u(x^2+\epsilon \gamma x+\frac{\gamma^2}{2})^{p^s+t-2i} h(x)\Big\},\vspace{-1mm}\end{equation*} which gives \vspace{-1mm}\begin{equation*}\text{ann}(\mathcal{I})=\left<(x^2+\epsilon \gamma x+\frac{\gamma^2}{2})^{p^s-i}-u(x^2+\epsilon \gamma x+\frac{\gamma^2}{2})^{p^s+t-2i} h(x)\right>.\vspace{-1mm}\end{equation*} Further, since $h(x)=\sum\limits_{j=0}^{i-t-1}A_j(x)(x^2+\epsilon \gamma x+\frac{\gamma^2}{2})^{j},$ by Lemma \ref{l13}, we get \small {\vspace{-1mm}\begin{equation*}\mathcal{I}^{\perp}=\Big< (\frac{\gamma^2}{2})^{p^s-i}(x^2+2\epsilon \gamma^{-1} x+\frac{2}{\gamma^{2}})^{p^s-i}-u \sum\limits_{j=0}^{i-t-1}A_j^*(x)x^{2i-2j-2t-\text{deg }A_j(x)} \\(\frac{\gamma^2}{2})^{p^s-2i+j+t}(x^2+2\epsilon \gamma^{-1} x+\frac{2}{\gamma^{2}})^{p^s-2i+j+t}\Big>.\vspace{-1mm}\end{equation*}}\normalsize

Next we assume that $\frac{p^s+t}{2} < i \leq p^s-1.$ In this case, \eqref{11} holds for $a=i-t,$ $b=0,$ $c=p^s-i$ and $g(x)=-h(x).$ This implies that $\text{ann}(\mathcal{I})=\left<(x^2+\epsilon \gamma x+\frac{\gamma^2}{2})^{i-t}-u h(x), u (x^2+\epsilon \gamma x+\frac{\gamma^2}{2})^{p^s-i}\right>.$ Further, since $h(x)=\sum\limits_{j=0}^{i-t-1}A_j(x)(x^2+\epsilon \gamma x+\frac{\gamma^2}{2})^{j}$ and  $\mathcal{I}^{\perp}=\text{ann}(\mathcal{I})^*,$ by Lemma \ref{l13}, we get $\mathcal{I}^{\perp}=\Big< (\frac{\gamma^2}{2})^{i-t}(x^2+2\epsilon \gamma^{-1} x+\frac{2}{\gamma^{2}})^{i-t}-u \sum\limits_{j=0}^{i-t-1}A_j^*(x)x^{2i-2j-2t-\text{deg }A_j(x)} (\frac{\gamma^2}{2})^{j}(x^2+2\epsilon \gamma^{-1} x+\frac{2}{\gamma^{2}})^{j}, u(x^2+2\epsilon \gamma^{-1} x+\frac{2}{\gamma^{2}})^{p^s-i}\Big>.$
\\(e) First let $h(x)=0$ so that $\mathcal{I}=\Big<(x^2+\epsilon \gamma x+\frac{\gamma^2}{2})^i,u(x^2+\epsilon \gamma x+\frac{\gamma^2}{2})^w \Big>.$ Here \vspace{-1mm}\begin{equation*}\text{ann}(\mathcal{I})=\Big<(x^2+\epsilon \gamma x+\frac{\gamma^2}{2})^{p^s-w}, u(x^2+\epsilon \gamma x+\frac{\gamma^2}{2})^{p^s-i} \Big>,\vspace{-1mm}\end{equation*} which, by Lemma \ref{l13}, implies that \vspace{-1mm}\begin{equation*}\mathcal{I}^{\perp}=\text{ann}(\mathcal{I})^*=\Big<(x^2+2\epsilon \gamma^{-1} x+\frac{2}{\gamma^{2}})^{p^s-w}, u(x^2+2\epsilon \gamma^{-1} x+\frac{2}{\gamma^{2}})^{p^s-i} \Big>.\vspace{-1mm}\end{equation*}

Next we suppose that $h(x)$ is a unit in $\mathsf{R}_{\epsilon \gamma}.$ By Proposition \ref{U}, we see that $U=\min\{i,p^s-i+t\}.$ As $w < U,$ we have $w < i$ and $w < p^s-i+t.$ In order to determine $\mathcal{I}^{\perp},$ we see that  $\text{ann}(\mathcal{I})$ is an ideal of $\mathsf{R}_{\epsilon \gamma}.$ So by Theorem \ref{t7}, we can write \vspace{-1mm}\begin{equation*}\text{ann}(\mathcal{I})=\left<(x^2+\epsilon \gamma x+\frac{\gamma^2}{2})^a+u(x^2+\epsilon \gamma x+\frac{\gamma^2}{2})^b g(x), u (x^2+\epsilon \gamma x+\frac{\gamma^2}{2})^c\right>,\vspace{-1mm}\end{equation*} where $g(x)$ is either 0 or a unit in $\mathsf{R}_{\epsilon\gamma},$ $c$ is the smallest non-negative integer satisfying \vspace{-1mm}\begin{equation*}u(x^2+\epsilon \gamma x+\frac{\gamma^2}{2})^{c}\in \Big< (x^2+\epsilon \gamma x+\frac{\gamma^2}{2})^a+u(x^2+\epsilon \gamma x+\frac{\gamma^2}{2})^b g(x)\Big>\vspace{-1mm}\end{equation*} and $a,b$ are integers satisfying $1 \leq a \leq p^s-1$ and $0 \leq b < c.$ This implies that  \begin{equation}\label{22} u(x^2+\epsilon \gamma x+\frac{\gamma^2}{2})^{i+c}= u(x^2+\epsilon \gamma x+\frac{\gamma^2}{2})^{a+w}=0 \text{ and }\end{equation} \begin{equation}\label{220} (x^2+\epsilon \gamma x+\frac{\gamma^2}{2})^{a+i}+u\left\{(x^2+\epsilon \gamma x+\frac{\gamma^2}{2})^{a+t}h(x)+(x^2+\epsilon \gamma x+\frac{\gamma^2}{2})^{i+b}g(x)\right\}=0.\end{equation} It is easy to see that \eqref{22} and\eqref{220} hold for $a=p^s-w,$ $b=p^s-w+t-i >0,$ $c=p^s-i$ and $g(x)=-h(x).$  This implies that \vspace{-1mm}\begin{equation*}\text{ann}(\mathcal{I})=\left<(x^2+\epsilon \gamma x+\frac{\gamma^2}{2})^{p^s-w}-u(x^2+\epsilon \gamma x+\frac{\gamma^2}{2})^{p^s+t-i-w} h(x), u (x^2+\epsilon \gamma x+\frac{\gamma^2}{2})^{p^s-i}\right>.\vspace{-1mm}\end{equation*}
Further, since $h(x)=\sum\limits_{j=0}^{w-t-1}A_j(x)(x^2+\epsilon \gamma x+\frac{\gamma^2}{2})^{j-t},$ using Lemma \ref{l13}, we get 

$\mathcal{I}^{\perp}=\text{ann}(\mathcal{I})^*=\Big< (\frac{\gamma^2}{2})^{p^s-w}(x^2+2\epsilon \gamma^{-1} x+\frac{2}{\gamma^{2}})^{p^s-w}-u \sum\limits_{j=0}^{w-t-1}A_j^*(x)x^{2i-2j-2t-\text{deg }A_j(x)} (\frac{\gamma^2}{2})^{p^s-i-w+j+t}$ \begin{flushright}$(x^2+2\epsilon \gamma^{-1} x+\frac{2}{\gamma^{2}})^{p^s-i-w+j+t},  u(x^2+2\epsilon \gamma^{-1} x+\frac{2}{\gamma^{2}})^{p^s-i}\Big>.$\end{flushright}
This completes the proof of the theorem. $\hfill \Box$\\
In the following corollary, we determine some isodual $\alpha$-constacyclic codes of length $4p^s$ over $R.$
\begin{cor} \label{c3}  \begin{enumerate}\item[(a)] The ideal $\left<u\right> \subseteq \mathfrak{R}_{\alpha}$ is an isodual $\alpha$-constacyclic code of length $4p^s$ over $R.$\item[(b)] For $1 \leq i,j \leq p^s-1,$ the ideal \vspace{-1mm}\begin{equation*}\Big<(x^2+\gamma x+\frac{\gamma^2}{2})^i,u(x^2+\gamma x+\frac{\gamma^2}{2})^{p^s-i}\Big>\oplus \Big<(x^2-\gamma x+\frac{\gamma^2}{2})^{j},u(x^2-\gamma x+\frac{\gamma^2}{2})^{p^s-j}\Big> \subseteq \mathfrak{R}_{\alpha}\vspace{-1mm}\end{equation*} is an isodual $\alpha$-constacyclic code of length $4p^s$ over $R.$   \end{enumerate}\end{cor}
\noindent \textbf{Proof.} Let $\mathcal{C}$ be an $\alpha$-constacyclic code of length $4p^s$ over $R.$ By Lemma \ref{l7}, we have $\mathcal{C}=\mathcal{C}_{1}\oplus \mathcal{C}_{2},$  where $\mathcal{C}_1,~\mathcal{C}_{2}$ are respectively the ideals of  rings $\mathsf{R}_{\gamma},~\mathsf{R}_{-\gamma}$ and as determined in Theorem \ref{t7}. For a code $\mathcal{C}$ to be isodual, we must have $|\mathcal{C}|=|\mathcal{C}^{\perp}|.$\\
(a) To prove this, let us take $\mathcal{C}_{1}=\left<u(x^2+\gamma x+\frac{\gamma}{2})^i \right>$ and $\mathcal{C}_{2}=\left<u(x^2-\gamma x+\frac{\gamma}{2})^j \right>$ for some integers $i$ and $j,$ where $0 \leq i,j \leq p^s-1.$ Now by Theorems \ref{t5} and \ref{tt6}, we see that  \vspace{-1mm}\begin{equation*}\mathcal{C}_{1}^{\perp}=\left<(x^2+2\gamma^{-1} x+\frac{2}{\gamma^2})^{p^s-i},u \right>, ~~~\mathcal{C}_{2}^{\perp}= \left<(x^2-2\gamma^{-1} x+\frac{2}{\gamma^2})^{p^s-j},u \right>,\vspace{-1mm}\end{equation*}  \vspace{-1mm}\begin{equation*} |\mathcal{C}_{1}|=p^{2m(p^s-i)},~~~|\mathcal{C}_{2}|=p^{2m(p^s-j)},~~|\mathcal{C}_{1}^{\perp}|=p^{2m(p^s+i)}\text {and } |\mathcal{C}_{2}^{\perp}|=p^{2m(p^s+j)}. \vspace{-1mm}\end{equation*} From this and using Lemmas \ref{l7} and \ref{d3}, we see that the code $\mathcal{C}$ is isodual if $|\mathcal{C}|=|\mathcal{C}_{1}||\mathcal{C}_{2}|=|\mathcal{C}_{1}^{\perp}||\mathcal{C}_{2}^{\perp}|=|\mathcal{C}^{\perp}|,$ which implies that $2p^s-i-j=2p^s+i+j.$ This holds only when $i+j=0,$ which gives $i=j=0.$ On the other hand, when $i=j=0,$ we have $\mathcal{C}_{1}=\left<u\right>,$ $\mathcal{C}_{2}=\left<u\right> ,$  $\mathcal{C}_{1}^{\perp}=\left<u\right>$ and $\mathcal{C}_{2}^{\perp}= \left<u\right>.$  This implies that $\mathcal{C}=\left<u\right> \subseteq \mathfrak{R}_{\alpha}$ and $\mathcal{C}^{\perp}=\left<u\right> \subseteq \mathfrak{R}_{\alpha^{-1}},$ which are trivially $R$-linearly equivalent.  \\
(b) To prove this, in view of Theorem \ref{t7}, let us take \vspace{-1mm}\begin{equation*}\mathcal{C}_{1}=\Big<(x^2+\gamma x+\frac{\gamma^2}{2})^{i}, u(x^2+\gamma x+\frac{\gamma^2}{2})^{w}\Big> \text{ and } \mathcal{C}_{2}= \Big<(x^2-\gamma x+\frac{\gamma^2}{2})^{j}, u(x^2-\gamma x+\frac{\gamma^2}{2})^{d}\Big>,\vspace{-1mm}\end{equation*} where $1 \leq i,j \leq p^s-1,$  $0 \leq w < i$ and $0 \leq d < j.$ By Theorems \ref{t5} and \ref{tt6}, we have \vspace{-1mm}\begin{equation*} |\mathcal{C}_1|=p^{2m(2p^s-i-w)}, ~~ |\mathcal{C}_{1}^{\perp}|=p^{2m(i+w)},~~\mathcal{C}_{1}^{\perp}=\Big<(x^2+2\gamma^{-1} x+\frac{2}{\gamma^2})^{p^s-w},u(x^2+2\gamma^{-1} x+\frac{2}{\gamma^2})^{p^s-i}\Big>,\vspace{-1mm}\end{equation*}\vspace{-1mm}\begin{equation*}   |\mathcal{C}_{2}|=p^{2m(2p^s-j-d)},~~ |\mathcal{C}_{2}^{\perp}|=p^{2m(j+d)}  \text{ and } \mathcal{C}_{2}^{\perp}= \Big<(x^2-2\gamma^{-1} x+\frac{2}{\gamma^2})^{p^s-d},u(x^2-2\gamma^{-1} x+\frac{2}{\gamma^2})^{p^s-j}\Big>.\vspace{-1mm}\end{equation*}  Now by Lemmas \ref{l7} and \ref{d3}, we see that the code $\mathcal{C}$ is isodual only if  $|\mathcal{C}|=|\mathcal{C}_{1}||\mathcal{C}_{2}|=|\mathcal{C}_{1}^{\perp}||\mathcal{C}_{2}^{\perp}|=|\mathcal{C}^{\perp}|,$ which implies that $i+w+j+d=2p^s.$ On the other hand, if $i+w=p^s$ and $j+d=p^s$ (so that $i+w+j+d=2p^s$ holds), we see that  \vspace{-1mm}\begin{equation*}  \mathcal{C}=\Big<(x^2+\gamma x+\frac{\gamma^2}{2})^{i}, u(x^2+\gamma x+\frac{\gamma^2}{2})^{p^s-i}\Big> \oplus \Big<(x^2-\gamma x+\frac{\gamma^2}{2})^{j}, u(x^2-\gamma x+\frac{\gamma^2}{2})^{p^s-j}\Big> \text{ and }\vspace{-1mm}\end{equation*} \vspace{-1mm}\begin{equation*}  \mathcal{C}^{\perp}=\Big<(x^2+2\gamma^{-1} x+\frac{2}{\gamma^2})^{i},u(x^2+2\gamma^{-1} x+\frac{2}{\gamma^2})^{p^s-i}\Big> \oplus \Big<(x^2-2\gamma^{-1} x+\frac{2}{\gamma^2})^{j}, u(x^2-2\gamma^{-1} x+\frac{2}{\gamma^2})^{p^s-j}\Big>,\vspace{-1mm}\end{equation*}  which are trivially $R$-linearly equivalent.

This completes the proof. $\hfill \Box$

\end{document}